%% file: main.tex
\documentclass[format=manuscript, screen=true, review=false]{acmart}

\AtBeginDocument{%
  }
\setcopyright{acmcopyright}
\copyrightyear{2024}
\acmYear{2024}
\acmDOI{XXXXXXX.XXXXXXX}

\usepackage{amssymb}
\usepackage{amsmath}
\usepackage{algorithm} 
\usepackage{algorithmic}
\usepackage{natbib}
\usepackage{bbm}
\usepackage{bm}
\usepackage{setspace}
\usepackage{multirow}
\hypersetup{colorlinks=true, linkcolor=blue, anchorcolor=blue, citecolor=red}
\usepackage{tikz}
\usepackage{graphicx}
\usetikzlibrary{shapes, arrows.meta, positioning}
\usepackage{tabularx}
\usepackage{booktabs}
\usepackage{multirow}
\usepackage{placeins}
\usepackage{amsmath}
\usepackage{graphicx}
\usepackage{booktabs}
\usepackage{fontawesome5}
\usepackage{booktabs}      
\usepackage[table]{xcolor} 
\usepackage{arydshln}

\usepackage{xcolor}
\usepackage[edges]{forest}

\usepackage{graphicx}
\usepackage{booktabs}
\usepackage{array}
\usepackage{adjustbox}

\usepackage{makecell}

\usepackage[utf8]{inputenc}
\usepackage{booktabs, multirow, tabularx}
\usepackage[table]{xcolor}
\usepackage{pifont}
\newcommand{\xmark}{\ding{55}} % ✗
\newcommand{\cmark}{\ding{51}} % ✓
\newcommand{\autocite}[1]{Author~et~al.~[\cite{#1}]} % 表格引用格式

\usepackage[globalcitecopy]{bibunits} % globalcitecopy 允许在不同部分引用同一文献
\defaultbibliographystyle{plain}       % 设置默认样式
\defaultbibliography{reference}         % 设置默认 bib 文件（不带 .bib 后缀）

\usetikzlibrary{arrows,shapes,positioning}
\usetikzlibrary{calc,decorations.markings}
\acmJournal{JACM}
\usepackage{paralist}

\begin{document}

\title{Document Parsing Unveiled: Techniques, Challenges, and Prospects for Structured Data Extraction} 

% \affiliation{%

\author{Qintong Zhang}
\authornote{Authors contributed equally to this research.}
% \authornote{The work is done during an internship at Shanghai Artificial Intelligence Laboratory.}
\affiliation{%
  \institution{Peking University}
  \city{Beijing}
  \country{China}
}
\email{qtzhang25@stu.pku.edu.cn}

\author{Bin Wang}
\authornotemark[1]
\affiliation{%
  \institution{Shanghai Artificial Intelligence Laboratory}
  \city{Shanghai}
  \country{China}
}
\email{wangbin@pjlab.org.cn}

\author{Victor Shea-Jay Huang}
\affiliation{%
  \institution{Beihang University}
  \city{Beijing}
  \country{China}
}
\email{jeix782@gmail.com}

\author{Junyuan Zhang}
\affiliation{%
  \institution{Shanghai Artificial Intelligence Laboratory}
  \city{Shanghai}
  \country{China}
}
\email{junyuanpk@gmail.com}

\author{Zhengren Wang}
\affiliation{%
  \institution{Peking University}
  \city{Beijing}
  \country{China}
}
\email{wzr@stu.pku.edu.cn}

\author{Hao Liang}
\affiliation{%
  \institution{Peking University}
  \city{Beijing}
  \country{China}
}
\email{hao.liang@stu.pku.edu.cn}

% \author{Shawn Wang}
% % \orcid{1234-5678-9012}
% % \author{G.K.M. Tobin}
% \affiliation{%
%   \city{Beijing}
%   \country{China}
% }
% \email{shawn.wang.academic@outlook.com}

% \author{Matthieu Lin}
% % \orcid{1234-5678-9012}
% % \author{G.K.M. Tobin}
% \affiliation{%
%   \city{Beijing}
%   \country{China}
% }
% \email{linmatthieu@gmail.com}

% \author{Bin Cui}
% \affiliation{%
%  \institution{Peking University}
%  \country{China}
% }
% \email{bin.cui@pku.edu.cn}

\author{Conghui He}

\authornote{Conghui He is the corresponding author.}
\affiliation{%
  %\streetaddress{}
  \institution{Shanghai Artificial Intelligence Laboratory}
  %\city{Shang Hai}
  \country{China}
}
\email{heconghui@pjlab.org.cn}
% \affiliation{%
%   \streetaddress{Center for Machine Learning Research}
%   \institution{Peking University}
%   \city{Beijing}
%   \country{China}
%   % \postcode{43017-6221}
% }
% \email{chrisallenming@gmail.com}

\author{Wentao Zhang}
% \authornotemark[3]
\affiliation{%
  %\streetaddress{Center for Machine Learning Research}
  \institution{Peking University}
  %\city{Beijing}
  \country{China}
}
\email{wentao.zhang@pku.edu.cn}

% \author{}
% \affiliation{%
%   \streetaddress{Institute of Advanced Technology}
%   \institution{University of Science and Technology of China}
%   \city{Hefei}
%   \country{China}
%   % \postcode{43017-6221}
% }
% \email{}
% \renewcommand{\shortauthors}{}
% \newcommand{\junyuan}[1]{\textcolor{red}{junyuan: #1}}

\renewcommand{\shortauthors}{Zhang et al.}

\begin{abstract}
Document parsing (DP) transforms unstructured or semi-structured documents into structured, machine-readable representations, enabling downstream applications such as knowledge base construction and retrieval-augmented generation (RAG). This survey provides a comprehensive and timely review of document parsing research. We propose a systematic taxonomy that organizes existing approaches into modular pipeline-based systems and unified models driven by Vision-Language Models (VLMs). We provide a detailed review of key components in pipeline systems, including layout analysis and the recognition of heterogeneous content such as text, tables, mathematical expressions, and visual elements, and then systematically track the evolution of specialized VLMs for document parsing. Additionally, we summarize widely adopted evaluation metrics and high-quality benchmarks that establish current standards for parsing quality. Finally, we discuss key open challenges, including robustness to complex layouts, reliability of VLM-based parsing, and inference efficiency, and outline directions for building more accurate and scalable document intelligence systems.

\end{abstract}

%    <concept>
%        <concept_id>10010147.10010257.10010258.10010262.10010277</concept_id>
%        <concept_desc>Computing methodologies~Transfer learning</concept_desc>
%        <concept_significance>500</concept_significance>
%        </concept>
%    <concept>
%        <concept_id>10010147.10010257.10010293</concept_id>
%        <concept_desc>Computing methodologies~Machine learning approaches</concept_desc>
%        <concept_significance>300</concept_significance>
%        </concept>
%        <concept>
%        <concept_id>10010147.10010178.10010205</concept_id>
%        <concept_desc>Computing methodologies~Search methodologies</concept_desc>
%        <concept_significance>300</concept_significance>
%        </concept>
%  </ccs2012>


% \ccsdesc[500]{Computing methodologies~Artificial intelligence~Natural language processing}
% \ccsdesc[500]{Computing methodologies~Artificial intelligence~Computer vision}


% \ccsdesc[500]{Computing methodologies~}
% \ccsdesc[300]{Computing methodologies~}
% \ccsdesc[300]{Computing methodologies~}


% \iffalse
% \received{XXX 2024}
% \received[revised]{XXX 2024}
% \received[accepted]{XXX 2024}
% \fi

\maketitle

% \input{section/sec1_Introduction}
% \input{section/sec2_Method}
% \input{section/sec_1_5_Definition_and_Relatedwork}
% \input{section/sec3_Document_Layout_Analysis}
% \input{section/sec4_Optical_Character_Recognition}
% \input{section/sec5_Mathematical_Formula_Detection_and_Recognition}
% \input{section/sec6_Table_Detection_and_Recognition}
% \input{section/sec7_special_elements_parsing}
% \input{section/sec8_Large_Language_Model_for_Document_Content_Extraction}
% \input{section/sec10_Discussion}
% % \input{section/sec12_conclusion}

% \bibliographystyle{plain}
% \bibliography{reference}

% \input{section/appendix}

\begin{bibunit}
    \input{section/sec1_Introduction}
    \input{section/sec_1_5_Definition_and_Relatedwork}

    \input{section/sec2_Method}
    \input{section/sec3_Document_Layout_Analysis}
    \input{section/sec4_Optical_Character_Recognition}
    \input{section/sec5_Mathematical_Formula_Detection_and_Recognition}

    \input{section/sec6_Table_Detection_and_Recognition}

    \input{section/sec7_special_elements_parsing}
    \input{section/sec8_Large_Language_Model_for_Document_Content_Extraction}
    \input{section/sec10_Discussion}
    \input{section/sec11_conclusion}
    \renewcommand{\bibname}{References} 
    \putbib
\end{bibunit}

\begin{bibunit}
    \input{section/appendix}

    \clearpage
    \renewcommand{\refname}{Appendix References}
    \putbib
\end{bibunit}

\end{document}

%% file: section/sec1_Introduction.tex
\section{Introduction}

As digital transformation accelerates, electronic documents have increasingly replaced paper as the primary medium for information exchange across industries, expanding both the diversity and complexity of document types and creating an urgent need for efficient systems to manage, analyze, and retrieve information~\cite{kerroumi2021visualwordgrid, fu2024ocrbench,yang2025cc}. Despite this transition, a large portion of historical archives, academic publications, and legal documents still exist in scanned or image-based formats, posing substantial challenges for downstream applications such as information extraction, semantic comprehension, and high-fidelity retrieval~\cite{subramani2020survey,baviskar2021efficient,xia2024docgenome,cui2026paddleocr15}.

\begin{figure}[!t]
\centering
\includegraphics[width=1.0\textwidth]{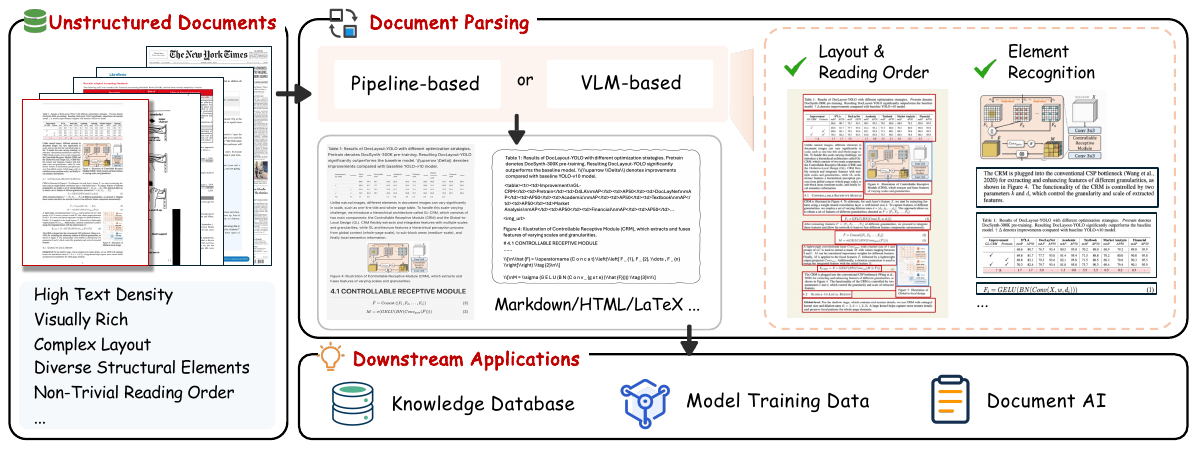}
\vspace{-5mm}
\caption{Overview of a document parsing pipeline: transforming unstructured pages into structured, machine-readable outputs, serving as a critical preprocessing step for document-centric downstream tasks. } 
\vspace{-5mm}
\label{fig1}
\end{figure}

To bridge this gap, Document Parsing (DP) has emerged as a fundamental technology in document intelligence systems~\cite{wei2024general, wang2024mineru}. As illustrated in Figure~\ref{fig1}, document parsing refers to the process of converting unstructured or semi-structured documents into structured, machine-readable representations~\cite{ouyang2024omnidocbench,zhang2024ocr}.
Unlike traditional Optical Character Recognition (OCR), which focuses primarily on transcribing textual content, document parsing aims to extract diverse document elements while preserving structural relationships and semantic organization, and often outputs structured formats such as Markdown, JSON, or XML for downstream integration~\cite{got2024improved}. Its importance has been further amplified by advances in Large Language Models (LLMs) and multimodal foundation models: structured document representations are central to building knowledge bases and training corpora for Retrieval-Augmented Generation (RAG)~\cite{lewis2020retrieval,gao2023retrieval,zhao2024retrieval}, and provide a foundation for knowledge grounding~\cite{wang2023docllm, wang2024mineru,zhao2024doclayout,wang2024cdm}, document-level reasoning and question answering~\cite{zhang2026docdancer,zhu2025doclens,wang2026agenticocr}.

Despite the rapid progress of deep learning–based document parsing tools, the literature still lacks a comprehensive and up-to-date synthesis that treats document parsing as an integrated, page-level problem. Existing surveys related to this area can generally be grouped into two categories.
The first category consists of single-task surveys that focus on specific components within the document processing pipeline, such as document layout analysis~\cite{binmakhashen2019document, kumar2025exploring}, OCR~\cite{islam2017survey, nguyen2021survey}, mathematical expression recognition~\cite{aggarwal2022survey}, table detection and structure recognition~\cite{salaheldin2024deep, prajapati2025table}, and chart understanding~\cite{davila2020chart,farahani2023automatic}. While valuable, they typically treat each task in isolation and provide limited guidance for modern document parsing systems where multiple components must be jointly modeled and integrated.
The second category includes broader surveys on document understanding systems~\cite{subramani2020survey,igorevna2022document}. These works review multiple stages, such as OCR, layout analysis, and downstream information extraction, but often emphasize traditional pipelines and were largely conducted before the recent surge of multimodal foundation models, thus overlooking the emerging paradigm of Vision-Language Models (VLMs) for unified document parsing and structured generation.

These limitations highlight the need for a holistic and up-to-date survey that systematically organizes the field from a unified perspective. In this work, we address these gaps by providing (i) a clear and comprehensive taxonomy covering the full document parsing workflow, (ii) a systematic review of both traditional modular pipelines and emerging unified document parsing models driven by VLMs, and (iii) a consolidated analysis of datasets, benchmarks, and evaluation protocols, together with an in-depth discussion of open challenges and future research directions. By bridging the gap between fragmented task-specific literature and modern unified parsing frameworks, this survey aims to provide a structured reference for researchers and practitioners working on next-generation document intelligence systems.

In this survey, we analyze advancements in document parsing from a holistic perspective. Our key contributions are summarized as follows:

\begin{itemize}
    \item \textbf{In-depth Analysis of Key Components.} We examine critical stages in the document parsing pipeline, including layout analysis and the recognition of multimodal elements such as text, tables, formulas, and charts.
    
    \item \textbf{Evolution of Specialized VLMs.} We track the development of VLMs for document parsing and analyze how they enable unified document understanding and structured generation.
    
    \item \textbf{Consolidation of Benchmarks and Metrics.} We summarize commonly used evaluation metrics and benchmark datasets for document parsing and its subtasks.
    
    \item \textbf{Identification of Challenges and Future Directions.} We discuss persistent challenges and outline future research directions.
\end{itemize}

The remainder of this paper is organized as follows. Section~\ref{sec:methodology} introduces our systematic taxonomy and paper selection strategy. Sections~\ref{sec:DLA},~\ref{sec:OCR},~\ref{sec:MER},~\ref{sec:table}, and~\ref{sec:other_elements} examine the core algorithms of modular pipeline systems, including layout analysis and element-specific recognition tasks. Section~\ref{sec:VLM} reviews the evolution of VLMs for document parsing. Section~\ref{sec:discussion} discusses current challenges and future research opportunities and Section~\ref{sec:conclusion} concludes the paper. We  summarizes commonly used evaluation metrics, benchmarks and datasets in Appendix~\ref{sec:eval} and ~\ref{sec:datasets}.

%% file: section/sec_1_5_Definition_and_Relatedwork.tex
\section{Problem Definition and Related Surveys}

\subsection{Problem Definition of Document Parsing}

Document parsing refers to the process of converting visually structured documents into machine-readable and semantically structured representations. Given a document page represented as an image or rendered page, the goal is to extract the document content and its structural organization, including textual elements, layout structures, and other visual components, and represent them in a structured format such as \texttt{JSON}, \texttt{HTML}, \texttt{LaTeX}, or \texttt{Markdown}.

Formally, let $D$ denote a document page drawn from the space of document images $\mathcal{D}$. A document typically contains heterogeneous elements, including text blocks, tables, mathematical expressions, charts, and other visual objects, arranged under a layout that encodes spatial and logical relationships. The objective of document parsing is to learn a mapping
\[
f: \mathcal{D} \rightarrow \mathcal{S},
\]
where $\mathcal{S}$ denotes the space of structured document representations. The output $S = f(D)$ captures both the document content and its structural organization, including element types, spatial relationships, and reading order.

Under this formulation, document parsing aims to recover a structured representation that faithfully preserves the semantic content and layout structure of the original document, enabling downstream applications such as information extraction, document retrieval, and knowledge base construction.

\subsection{Comparison with Existing Surveys}

Based on their scope and objectives, existing surveys related to document parsing can generally be categorized into two groups: surveys focusing on individual tasks and surveys covering document parsing systems. Table~\ref{tab:survey_comparison} summarizes representative existing surveys and highlights the differences between them and our survey.

\paragraph{Surveys for Single Tasks.} 
Early research on document analysis did not yet form a unified framework for document parsing. Instead, researchers typically focused on improving the performance of individual subtasks within the document processing pipeline. Consequently, many surveys from this line of work concentrate on a single component of document parsing.
Among these tasks, Document Layout Analysis(DLA) has long been regarded as a fundamental preprocessing step for document understanding. For example, \cite{binmakhashen2019document} provides a comprehensive overview of early rule-based and vision-based approaches for DLA. More recently, \cite{kumar2025exploring} further summarizes modern deep learning--based layout analysis methods, together with representative datasets and emerging challenges associated with increasingly diverse document layouts.
A number of surveys focus on recognizing key document elements. For instance, several surveys review progress in text-related tasks, including Optical Character Recognition (OCR), post-OCR processing, and evaluation protocols~\cite{islam2017survey,nguyen2021survey,neudecker2021survey}. Other surveys investigate the recognition of specialized structured content such as mathematical expressions~\cite{aggarwal2022survey,sakshi2024machine} and table detection or table structure recognition~\cite{salaheldin2024deep,prajapati2025table}. 
Beyond text-centric elements, documents also contain rich visual information that conveys essential knowledge. Surveys such as~\cite{farahani2023automatic,huang2024pixels} review the automatic understanding of charts and graphical visualizations, covering tasks such as chart classification, data extraction, and chart question answering. In addition, chemical structures frequently appear in scientific documents, particularly in chemistry and biomedical literature. The survey by~\cite{musazade2022review} summarizes recent techniques and datasets for Optical Chemical Structure Recognition (OCSR).
Although these surveys provide valuable insights into specific document elements, they typically target isolated parsing or extraction tasks. As a result, their scope is limited when considering modern end-to-end document parsing systems, where multiple components must be jointly modeled and optimized.

\begin{table}[!t]
    \centering
    \small
    \setlength{\tabcolsep}{4pt}
    \renewcommand{\arraystretch}{0.95}
    \caption{Comparison of representative surveys related to document parsing. The table contrasts prior surveys in terms of task coverage across layout analysis, text, mathematical expressions, tables, and visual elements, and indicates whether each survey discusses VLM-based document parsing.}
    
    \begin{tabular}{l c c c c c c c}
        \toprule
        \multirow{2}{*}{\textbf{Survey}} & \multirow{2}{*}{\textbf{Year}} & \multicolumn{5}{c}{\textbf{Single Task Coverage}} & \multirow{2}{*}{\textbf{\makecell{VLM \\ for DP}}} \\
        \cmidrule(lr){3-7}
        & & \makecell{Layout \\ Analysis} & Text & \makecell{Mathematical \\ Expression} & Table & \makecell{Visual \\ Elements} & \\
        \midrule
        \rowcolor[gray]{0.95} \multicolumn{8}{l}{\textit{Survey for Single Task}} \\
        ~\cite{binmakhashen2019document} & 2019 & \cmark & \xmark & \xmark & \xmark & \xmark & \xmark \\
        \cite{kumar2025exploring} & 2025 & \cmark & \xmark & \xmark & \xmark & \xmark & \xmark \\
        \cite{islam2017survey} & 2017 & \xmark & \cmark & \xmark & \xmark & \xmark & \xmark \\
        \cite{nguyen2021survey} & 2021 & \xmark & \cmark & \xmark & \xmark & \xmark & \xmark \\
        \cite{salaheldin2024deep} & 2024 & \xmark & \xmark & \xmark & \cmark & \xmark & \xmark \\
        \cite{prajapati2025table} & 2025 & \xmark & \xmark & \xmark & \cmark & \xmark & \xmark \\
        \cite{aggarwal2022survey} & 2022 & \xmark & \xmark & \cmark & \xmark & \xmark & \xmark \\
        \cite{farahani2023automatic} & 2023 & \xmark & \xmark & \xmark & \xmark & Chart & \xmark \\
        \cite{musazade2022review} & 2022 & \xmark & \xmark & \xmark & \xmark & Chemical & \xmark \\
        \midrule
        \rowcolor[gray]{0.95} \multicolumn{8}{l}{\textit{Survey for Document Parsing System}} \\
        \cite{subramani2020survey} & 2020 & \cmark & \cmark & \xmark & \xmark & \xmark & \xmark \\
        \cite{igorevna2022document} & 2022 & \cmark & \cmark & \xmark & \cmark & \xmark & \xmark \\
        \textbf{Ours} & \textbf{2026} & \cmark & \cmark & \cmark & \cmark & Chart/Chemical & \cmark \\
        \bottomrule
    \end{tabular}
    \vspace{-5mm}
    \label{tab:survey_comparison}
\end{table}

\paragraph{Surveys for Document Parsing Systems.}
Compared with task-specific surveys, only a limited number of works attempt to provide a more integrated overview of document understanding systems. Representative examples include the surveys by \cite{subramani2020survey} and \cite{igorevna2022document}, which discuss several key stages such as OCR, layout analysis, and downstream information extraction. However, these works often treat document understanding primarily as a combination of layout analysis and OCR, resulting in taxonomies that are insufficient to cover the full document parsing workflow and the diverse set of structured elements present in modern documents.
Furthermore, these surveys were conducted prior to the rapid emergence of multimodal foundation models. Consequently, they largely overlook recent advances in VLMs that enable end-to-end document parsing and structured generation from raw document inputs.
To address these limitations, this survey makes the following contributions: 
(i) we propose a clear and comprehensive taxonomy that covers the entire document parsing workflow, including both textual and visual elements; 
(ii) we systematically review both traditional modular pipelines and recent VLM-based document parsing models, together with a detailed consolidation of benchmark datasets and evaluation protocols; and 
(iii) we present a structured analysis of the historical evolution of document parsing techniques, highlighting key challenges, emerging research trends, and open problems for future research.

%% file: section/sec2_Method.tex
\section{Methodology and Taxonomy}
\label{sec:methodology}

To provide a rigorous and theoretically grounded overview of document parsing research, this section outlines the literature retrieval and selection protocol and introduces the principles underlying the proposed taxonomy. 
By combining systematic retrieval with conceptually grounded categorization, we aim to ensure both comprehensive empirical coverage and a coherent organization of the evolving document parsing landscape.

\subsection{Literature Selection Strategy}

The literature review follows a structured process inspired by systematic survey methodologies in computing research, including identification, deduplication, title/abstract screening, and full-text eligibility assessment.

\subsubsection{Search Scope and Databases}

To ensure broad coverage, we conducted searches across multiple academic databases and scholarly search engines. 
The primary sources include the ACM Digital Library\footnote{\url{https://dl.acm.org/}} and IEEE Xplore\footnote{\url{https://ieeexplore.ieee.org/}}, which index major venues in document analysis and computer vision. 
Web of Science\footnote{\url{https://www.webofscience.com/}} and Scopus\footnote{\url{https://www.scopus.com/}} were used to capture interdisciplinary publications across information retrieval and machine learning. 
Google Scholar\footnote{\url{https://scholar.google.com/}} complements these databases with broad cross-domain coverage, while arXiv\footnote{\url{https://arxiv.org/}} provides access to recent advances in multimodal and vision-language modeling. 
Together, these sources cover research across document analysis, computer vision, natural language processing, and multimodal learning.

\subsubsection{Search Keywords and Query Design}

Search queries were constructed by combining keywords describing document parsing tasks with those reflecting modern modeling paradigms. 
To reduce bias toward specific approaches, retrieval was organized into two stages capturing both task-centric and technology-driven perspectives.

The first stage focused on core document parsing tasks without restricting methodology, using terms such as ``document parsing'', ``layout analysis'', ``OCR'', ``table recognition'', ``formula recognition'', ``chart parsing'', ``OCSR'', and ``document structure extraction''. 
These keywords correspond to the fundamental sub-tasks of document parsing and ensure coverage of classical approaches, including rule-based and early machine learning methods.

The second stage targeted recent advances driven by deep learning and multimodal modeling. 
Keywords such as ``encoder-decoder document analysis'', ``vision-language model for document parsing'', and ``multimodal document understanding'' were used to capture developments associated with Transformer architectures, large-scale pretraining, and the integration of vision and language models. 
These queries reflect the growing role of unified and generative modeling in document parsing.

Across databases, queries were formulated using Boolean combinations of task and method terms.
The search covers publications from approximately 1995 to 2026, capturing both early foundational work and recent deep learning–based advances.

\subsubsection{Screening and Eligibility}

The retrieved records were filtered through a multi-stage process, as illustrated in Figure~\ref{fig:selection_flow}. 
The initial search yielded approximately $N \approx 860$ records, which were reduced to $N \approx 680$ after removing duplicates. 
Title and abstract screening excluded irrelevant studies and application-focused work without methodological contributions, leaving approximately $N \approx 429$ papers.

The remaining papers were assessed through full-text review to ensure methodological relevance and experimental rigor. 
Studies were included if they proposed or evaluated methods for document parsing or closely related sub-tasks and provided sufficient technical detail and empirical validation. 
Works lacking experimental evaluation, containing only high-level system descriptions, or focusing solely on application deployment were excluded. 
After this process, approximately $N = 230$ papers were retained for detailed analysis.

\begin{figure}[h]
\centering
\small
\fbox{Identification ($N \approx 860$) $\rightarrow$ Screening ($N \approx 680$) $\rightarrow$ Eligibility ($N \approx 429$) $\rightarrow$ Final Inclusion ($N = 230$)}
\caption{Systematic literature screening workflow. The figure summarizes the staged reduction from initial retrieval to the final set of papers analyzed in this survey, including duplicate removal, abstract screening, and full-text eligibility assessment.}
\label{fig:selection_flow}
\end{figure}

\definecolor{lightcoral}{rgb}{0.94, 0.5, 0.5}
\definecolor{lightgreen}{rgb}{0.56, 0.93, 0.56}
\definecolor{harvestgold}{rgb}{0.98, 0.85, 0.40}
\definecolor{brightlavender}{rgb}{0.75, 0.58, 0.89}
\definecolor{capri}{rgb}{0.0, 0.75, 1.0}
\definecolor{carminepink}{rgb}{0.92, 0.3, 0.26}
\definecolor{celadon}{rgb}{0.67, 0.88, 0.69}
\definecolor{darkpastelgreen}{rgb}{0.01, 0.75, 0.24}
\definecolor{hidden-draw}{RGB}{205, 44, 36}

\tikzstyle{my-box}=[
    rectangle,
    draw=hidden-draw,
    rounded corners,
    text opacity=1,
    minimum height=1.5em,
    minimum width=5em,
    inner sep=2pt,
    align=left,
    fill opacity=.5,
]
\tikzstyle{first_leaf}=[my-box, minimum height=1.5em,
    fill=harvestgold!20, text=black, align=left,font=\scriptsize,
    inner xsep=2pt,
    inner ysep=4pt,
]
\tikzstyle{second_leaf}=[my-box, minimum height=1.5em,
    fill=cyan!20, text=black, align=left,font=\scriptsize,
    inner xsep=2pt,
    inner ysep=4pt,
]
\tikzstyle{third_leaf}=[my-box, minimum height=1.5em,
    fill=lightgreen!20, text=black, align=left,font=\scriptsize,
    inner xsep=2pt,
    inner ysep=4pt,
]
\tikzstyle{fourth_leaf}=[my-box, minimum height=1.5em,
    fill=brightlavender!20, text=black, align=left,font=\scriptsize,
    inner xsep=2pt,
    inner ysep=4pt,
]
\tikzstyle{fifth_leaf}=[my-box, minimum height=1.5em,
    fill=orange!20, text=black, align=left,font=\scriptsize,
    inner xsep=2pt,
    inner ysep=4pt,
]
\tikzstyle{sixth_leaf}=[my-box, minimum height=1.5em,
    fill=red!20, text=black, align=left,font=\scriptsize,
    inner xsep=2pt,
    inner ysep=4pt,
]
\tikzstyle{seventh_leaf}=[my-box, minimum height=1.5em,
    fill=teal!20, text=black, align=left,font=\scriptsize,
    inner xsep=2pt,
    inner ysep=4pt,
]
\begin{figure*}[t]
    \centering
    \resizebox{\textwidth}{!}{
        \begin{forest}
            forked edges,
            for tree={
                grow=east,
                reversed=true,
                anchor=base west,
                parent anchor=east,
                child anchor=west,
                base=left,
                font=\scriptsize,
                rectangle,
                draw=hidden-draw,
                rounded corners,
                align=left,
                minimum width=3em,
                edge+={darkgray, line width=1pt},
                s sep=3pt,
                inner xsep=2pt,
                inner ysep=3pt,
                ver/.style={rotate=90, child anchor=north, parent anchor=south, anchor=center},
            },
            where n children=0{
                align=left,
            }{},
            where level=1{text width=5.0em,font=\scriptsize, align=left}{},
            where level=2{text width=9.8em,font=\scriptsize,align=left}{},
            where level=3{text width=10.5em,font=\scriptsize,align=left}{},
            where level=4{text width=5.0em,font=\scriptsize,}{},
            [
                Document Parsing, ver, color=carminepink!100, fill=carminepink!15, text=black
                [
                    Modular Pipeline-Based Systems, color=cyan!100, fill=cyan!100, text=black, text width=11em
                        [
                        Document Layout Analysis, color=brightlavender!100, fill=brightlavender!60, text=black
                        [
                            Visual-based Layout Detection, color=brightlavender!100, fill=brightlavender!40, text=black
                            [
                                {
                                    CNN-based Methods~\cite{oliveira2017fast,zhao2024doclayout,qi2025yolo,sun2025pp}\\
                                    Transformer-based Methods~\cite{bao2021beit,li2022dit,shehzadi2025docsemi}
                                }, fourth_leaf, text width=18em
                            ]
                        ]
                        [
                            Multimodal Layout Understanding, color=brightlavender!100, fill=brightlavender!40, text=black
                            [
                                {
                                    Pre-trained Models (LayoutLM v1-v3)~\cite{xu2020layoutlm,xu2020layoutlmv2,huang2022layoutlmv3}\\
                                    Grid \& Graph Representations~\cite{katti2018chargrid,denk2019bertgrid,wang2023graphical}\\
                                    LLM-based \& Retrieval~\cite{zhu2025simple,yan2026beyond}
                                }, fourth_leaf, text width=18em
                            ]
                        ]
                        [
                             Trends \& Frontiers, color=brightlavender!100, fill=brightlavender!40, text=black
                            [
                                {
                                    Dataset Diversity \& Fine-grained ~\cite{cheng2023m6doc}, \\
                                    Semi-supervised Learning ~\cite{shehzadi2025docsemi}
                                    % \\ Real-world Robustness \& Large-scale Construction
                                }, fourth_leaf, text width=18em
                            ]
                        ]
                    ]
[
                        Optical Character Recognition, color=harvestgold!100, fill=harvestgold!60, text=black
                        [
                            Foundations and Classical Pipelines, color=harvestgold!100, fill=harvestgold!40, text=black
                            [
                                {
                                    Text Detection~\cite{liao2017textboxes,zhou2017east,deng2018pixellink,baek2019character}\\
                                    Text Recognition~\cite{shi2016end,cheng2018aon,li2023trocr}\\
                                    End-to-End Text Spotting~\cite{liu2020abcnet,huang2022swintextspotter}
                                }, first_leaf, text width=18em
                            ]
                        ]
                        % [
                        %     Transition to Multimodal Intelligence, color=harvestgold!100, fill=harvestgold!40, text=black
                        %     [
                        %         {
                        %             General-purpose VLMs Paradigm Shift\\
                        %             Robustness in Complex/Distorted Scenes\\
                        %             OCR as a byproduct of Multimodal Alignment
                        %         }, first_leaf, text width=18em
                        %     ]
                        % ]
                        [
                            Trends \& Frontiers,, color=harvestgold!100, fill=harvestgold!40, text=black
                            [
                                {
                                    Vertical Scenarios~\cite{pandey2025survey,lu2025deepad},
                                    Internal Mechanism~\cite{baek2025large}, \\Reliability~\cite{zhang2025consensus}
                                }, first_leaf, text width=18em
                            ]
                        ]
                    ]
                    [
                        Mathematical Expression \\Detection and Recognition, color=orange!100, fill=orange!60, text=black
                        [
                            Mathematical Expression Detection, 
                            color=orange!100, fill=orange!40, text=black
                            [
                                {
                                    Object Detection~\cite{gao2017deep,yi2017cnn,li2018page,mali2020scanssd,younas2019ffd,nguyen2021ds}\\
                                    Layout-aware FormulaDet~\cite{hu2024mathematical}
                                    % Integration with Unified DLA Systems
                                }, fifth_leaf, text width=18em
                            ]
                        ]
                        [
                            Mathematical Expression \\Recognition, color=orange!100, fill=orange!40, text=black
                            [
                                {
                                    Sequence-based~\cite{zhao2021handwritten,zhao2022comer}\\
                                    Structure-aware~\cite{zhu2025tamer, zhang2025ssan,li2022counting, xie2025tan}\\
                                    Self-scale Alignment ~\cite{yang2025autoscaler}
                                }, fifth_leaf, text width=18em
                            ]
                        ]
                        [
                            Trends \& Frontiers, color=orange!100, fill=orange!40, text=black
                            [
                                {
                                    Large Datasets ~\cite{gervais2025mathwriting,wang2024unimernet} , Efficiency~\cite{liu2025pp,du2025unirec}, \\
                                    VLM-based~\cite{li2025uni,zhong2025doctron,guo2025hie}
                                }, fifth_leaf, text width=18em
                            ]
                        ]
                    ]
                    [
                        Table Detection and Recognition, color=teal!100, fill=teal!60, text=black
                        [
                            Table Detection, color=teal!100, fill=teal!40, text=black
                            [
                                {
                                    Object Detection~\cite{hao2016table,gilani2017table,schreiber2017deepdesrt,siddiqui2018decnt,huang2019yolo}\\
                                    Advanced Sparse Detection~\cite{xiao2023table}
                                }, seventh_leaf, text width=18em
                            ]
                        ]
                        [
                            Table Structure Recognition, color=teal!100, fill=teal!40, text=black
                            [
                                {
                                    Top-down~\cite{siddiqui2019deeptabstr,zou2020deep,guo2022trust,wang2023robust,khan2019table}\\
                                    Bottom-up~\cite{prasad2020cascadetabnet,long2021parsing,chi2019complicated,qasim2019rethinking}\\
                                    Feature Backbones~\cite{arik2021tabnet,huang2020tabtransformer}
                                }, seventh_leaf, text width=18em
                            ]
                        ]
                        [
                            Trends \& Frontiers , color=teal!100, fill=teal!40, text=black
                            [
                                {
                                    Encoder–Decoder \& MASTER~\cite{deng2019challenges,zhong2020image,ye2021pingan}\\
                                    Unified Frameworks~\cite{yang2025alignment,peng2024unitable}\\
                                    VLM-based \& Self-supervised~\cite{zhang2025trivia,wang2025transtab}
                                }, seventh_leaf, text width=18em
                            ]
                        ]
                    ]
                    [
                        Visual Element Parsing, color=red!100, fill=red!60, text=black
                        [
                            Chart Parsing, color=red!100, fill=red!40, text=black
                            [
                                {
                                    Pipeline-based~\cite{dhote2024swin,shaheen2024c2f,xu2024empowering,qiao2023structure,mustafa2023charteye}\\
                                    End-to-End Multimodal Generation~\cite{meng2024chartassistant,zhang2024tinychart,xu2024chartmoe}\\
                                    Chart-to-Code~\cite{zhao2025vincicoder,tang2025charts}
                                }, sixth_leaf, text width=18em
                            ]
                        ]
                        [
                            Optical Chemical Structure Recognition, color=red!100, fill=red!40, text=black, text width=11.5em
                            [
                                {
                                    Rule-based Graph Reconstruction~\cite{filippov2009optical,zimmermann2011chemical}\\
                                    Deep Sequence Generation~\cite{xu2022swinocsr,lin2024mpocsr,oldenhof2024atom}\\
                                    MLLMs \& Semantic Optimization~\cite{fang2025molparser,li2025chemvlm,zhao2025tinychemvl,zhang2025molsight}
                                }, sixth_leaf, text width=17em
                            ]
                        ]
                    ]
                ]
[
                    VLMs for Document Parsing, color=lightgreen!100, fill=lightgreen!100, text=black, text width=11em
                    [
                        General-Purpose VLMs, color=lightgreen!100, fill=lightgreen!60, text=black
                        [
                            {   
                                Open Sources VLMs~\cite{lu2024deepseek, wang2024qwen2, qwen3technicalreport, Qwen35, zhu2025internvl3, wang2025internvl35}, 
                                Proprietary VLMs~\cite{hurst2024gpt4o, singh2025gpt5, team2025kimi, kimik25, google2025gemini3pro, Claude}
                            }, third_leaf, text width=30em
                        ]
                    ]
                    [
                        Specialized VLMs, color=lightgreen!100, fill=lightgreen!60, text=black
                        [
                            End-to-End Parsing, color=lightgreen!100, fill=lightgreen!60, text=black
                            [
                                {
                                    % Nougat~\cite{blecher2023nougat},Vary~\cite{wei2025vary},
                                    GOT-OCR2.0~\cite{wei2024general},
                                    DeepSeek-OCR~\cite{wei2025deepseek,wei2026deepseek2}, 
                                    SmolDocling~\cite{nassar2025smoldocling}, \\
                                    UniRec~\cite{du2025unirec}, HunyuanOCR~\cite{team2025hunyuanocr}
                                    Logics-parsing~\cite{chen2025logics},\\
                                    olmOCR~\cite{poznanski2025olmocr1, poznanski2025olmocr},
                                    dots.ocr~\cite{li2025dots}, FireRed-OCR~\cite{wu2026firered}...
                                }, third_leaf, text width=18em
                            ]     
                        ]
                        [
                            Multi-Stage Parsing, color=lightgreen!100, fill=lightgreen!60, text=black
                            [
                                {
                                    Decoupled Inference (MinerU, GLM-OCR)~\cite{niu2025mineru2,duan2026glm}\\
                                    SRR Paradigm (MonkeyOCR v1/v1.5)~\cite{li2025monkeyocr,zhang2025monkeyocr}\\
                                    Robustness to Distortions (PaddleOCR-VL)~\cite{cui2025paddleocr,cui2026paddleocr15}
                                }, third_leaf, text width=18em
                            ]  
                        ]
                        [
                            Trends \& Frontiers, color=lightgreen!100, fill=lightgreen!60, text=black
                            [
                                {
                                    Hallucination Mitigation~\cite{he2025seeing,chen2025dianjin}\\
                                    Post-OCR Correction \& RL Improvement~\cite{shim2025revise,wang2025infinity}\\
                                    Fine-grained Quality Assessment~\cite{zhang2025docr,tan2025ocr,zhong2026ocrverse}
                                }, third_leaf, text width=18em                                            
                            ]
                        ]
                    ]
                ]
            ]
        \end{forest}
    }
    \caption{Overview of the survey taxonomy for document parsing.}
    \label{fig1:overview_table}
    \vspace{-3mm}
\end{figure*}

\subsection{Principles for Taxonomy Construction}

Before introducing the taxonomy used in this survey, we first outline the principles guiding the classification of document parsing approaches. Document parsing spans a broad range of tasks, datasets, and modeling paradigms, which makes the design of a coherent taxonomy inherently challenging. An effective framework should therefore emphasize fundamental methodological distinctions while remaining sufficiently general to accommodate diverse problem settings.

A primary consideration is architectural distinguishability. Methods should be grouped according to their underlying system organization rather than superficial implementation differences. In document parsing, architectural design largely determines how visual features, textual content, and structural representations are integrated.

Task generality is another key factor. Document parsing comprises multiple sub-problems, including layout analysis, optical character recognition, table understanding, mathematical expression recognition, and chart interpretation. A meaningful taxonomy should therefore apply across these heterogeneous tasks rather than being tailored to a specific application.

The taxonomy should also reflect the historical evolution of the field. Document analysis has progressed from rule-based and heuristic systems to statistical learning pipelines, and more recently to unified multimodal architectures. Capturing these transitions helps contextualize current approaches and clarifies how modeling assumptions have shifted over time.

Finally, the classification should align with common evaluation practices, datasets, and supervision strategies. Methods that share similar training objectives, supervision signals, and evaluation protocols often exhibit consistent design patterns and can be naturally grouped together.

Guided by these principles, the taxonomy presented in this survey primarily organizes document parsing methods according to their architectural structure.

\subsection{Architectural Principles of Classification}
\label{principles}

Document parsing can be formalized as a structured transformation problem that converts visually heterogeneous documents into machine-readable representations while preserving semantic hierarchy and reading order. Accordingly, the taxonomy adopted in this survey distinguishes between two dominant paradigms: \emph{modular pipeline-based systems} and \emph{unified vision-language models}.

Modular pipeline-based systems follow an explicit task decomposition strategy. 
Document parsing is divided into a sequence of intermediate sub-tasks such as layout detection, optical character recognition, table structure extraction, mathematical expression parsing, and visual element analysis. 
Each component is optimized separately with task-specific supervision and objective functions, while intermediate representations—such as bounding boxes, OCR tokens, or structural graphs—are propagated between stages. 
This paradigm emphasizes modularity, interpretability, and controllable optimization, although it often requires carefully designed task boundaries and coordination across heterogeneous modules.

In contrast, unified vision-language models approach document parsing as an end-to-end multimodal learning problem. 
Rather than relying on explicitly defined intermediate pipelines, these models learn shared visual and textual representations and directly generate structured outputs. 
Typically implemented using large Transformer-based architectures, they benefit from large-scale multimodal pretraining and parameter sharing, enabling stronger cross-task generalization. 
This paradigm reflects a broader shift in artificial intelligence from staged processing pipelines toward unified representation learning and generative modeling.

The distinction between these two paradigms forms the primary structural axis of this survey. 

\subsection{Hierarchical Organization of the Taxonomy}
\label{taxonomy}
Based on the above architectural principles, the subsequent sections are organized into \textbf{Modular Pipeline-Based Systems} and \textbf{VLMs for Document Parsing}. The complete classification system  are shown in Figure~\ref{fig3}.

Within the modular track, methods are analyzed according to functional components, including layout analysis, optical character recognition, mathematical expression processing, table parsing, and visual element parsing (e.g., chart parsing and OCSR). 
For each component, representative approaches are presented in roughly chronological order to reflect the methodological evolution from rule-based and statistical models to deep neural architectures.

Within the unified vision-language track, approaches are primarily organized according to their temporal development, highlighting the progressive shift from early encoder-decoder frameworks toward large-scale multimodal foundation models. 
In particular, we distinguish between two technical trajectories: models that perform fully end-to-end structured generation without explicit intermediate supervision, and models that retain certain multi-stage reasoning or intermediate decoding strategies within a unified parameter space. 
This distinction reflects different design choices regarding structural decomposition and optimization granularity under the unified modeling paradigm.

Across both tracks, methods are further examined through the lens of modeling architecture and learning strategy, and are discussed in developmental order to expose architectural transitions and emerging trends. 
Finally, we synthesize commonly adopted evaluation metrics, benchmark protocols, and publicly available datasets to provide a consolidated view of experimental practices in document parsing research.

By organizing the survey according to architectural paradigm, functional specialization, and temporal evolution, the proposed taxonomy maintains conceptual coherence while remaining flexible enough to accommodate emerging document foundation models.

%% file: section/sec3_Document_Layout_Analysis.tex
\section{Document Layout Analysis}
\label{sec:DLA}

Document layout analysis (DLA) aims to detect and classify structural elements in documents, such as text blocks, tables, figures, and formulas, and determine their spatial relationships. As a fundamental step in document parsing pipelines, DLA converts unstructured document images into structured representations that can support downstream content recognition.

\begin{figure}
\centering
\includegraphics[width=0.965\textwidth]{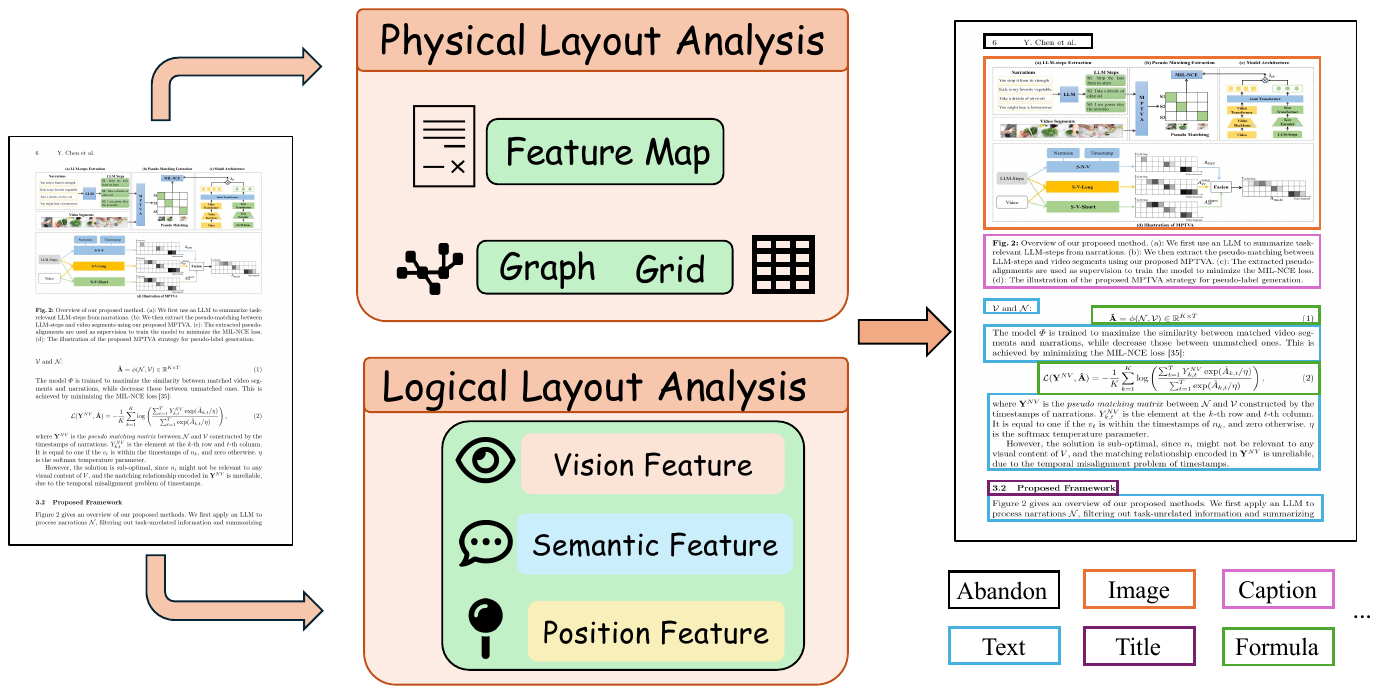}
% \vspace{-5mm}
\caption{Overview of document layout analysis. The figure summarizes the progression from page-level visual input to the detection, classification, and structural interpretation of layout elements such as text blocks, figures, tables, and formulas.} 
\label{fig3}
\vspace{-5mm}
\end{figure}

Early research on DLA mainly relied on rule-based and heuristic methods developed in the 1990s, which used handcrafted features and projection-based segmentation strategies to detect document regions. While these methods were effective for simple and well-formatted documents, they struggled to generalize to complex layouts and diverse document types. With the rapid development of deep learning and large-scale document datasets, DLA has evolved into a data-driven task and has become a core component of modern document intelligence systems. Figure~\ref{fig3} provides an overview of the main stages involved in document layout analysis.

Recent studies indicate that the development of DLA since 2020 has mainly progressed along three directions: (1) visual layout detection using deep neural networks, (2) multimodal layout understanding that integrates textual and semantic information, and (3) scalable and efficient learning paradigms designed for real-world document scenarios.

\subsection{Visual-based Layout Detection}

Most DLA approaches treat document pages as images and detect layout elements using visual features. With the advancement of deep learning, object detection and segmentation models have become the dominant paradigm for layout detection.

\subsubsection{CNN-based Methods}

Convolutional neural networks (CNNs) were among the earliest deep learning models applied to document layout analysis. Object detection frameworks such as Faster R-CNN and Mask R-CNN have been widely adopted to detect layout elements including text blocks, tables, and figures \cite{oliveira2017fast}. These methods significantly improved detection accuracy compared to rule-based approaches by learning hierarchical visual representations directly from data.
Relative to earlier heuristic pipelines, CNN-based methods offer substantially better adaptability to layout variation. Their locality-biased feature extraction, however, can make them less effective when long-range structural dependencies are critical.
In recent years, the YOLO family of models has gained popularity due to its high efficiency and real-time performance. For example, DocLayout-YOLO \cite{zhao2024doclayout} introduces a Global-to-Local Controllable Receptive Module to improve multi-scale layout detection while maintaining fast inference speed. Building upon this line of research, YOLO-DLA \cite{qi2025yolo} further investigates the challenges of multi-scale document layouts, where macro-scale elements (e.g., text, tables, figures) coexist with micro-scale elements such as headings, captions, and formulas. To address the imbalance in detecting elements of different scales, YOLO-DLA introduces Kernel Weighting Convolution and a scale-aware curriculum learning strategy that progressively trains models from macro to micro elements. This approach significantly improves detection performance for small layout components that are often overlooked in previous models.
For large-scale deployment, these models are generally more attractive than two-stage detectors because of their lower latency. Their gains in efficiency may still come at the cost of reduced robustness on densely packed or semantically ambiguous layouts.
Another recent effort focusing on practical deployment is PP-DocLayout \cite{sun2025pp}, which proposes a unified layout detection framework capable of processing diverse document types such as academic papers, books, magazines, and exam papers. The model supports 23 layout categories and achieves real-time inference speeds exceeding 120 pages per second, demonstrating the feasibility of large-scale document data construction for training modern document AI systems.

\subsubsection{Transformer-based Methods}

While CNN-based models perform well in capturing local visual features, they often struggle to model long-range dependencies across document regions. Transformer-based architectures address this limitation by using self-attention mechanisms to capture global contextual relationships within a document page.
Vision Transformer-based models such as BEiT \cite{bao2021beit} and the Document Image Transformer (DiT) \cite{li2022dit} apply patch-based representations to document images, enabling models to learn global layout patterns. These methods have demonstrated strong performance on several document analysis tasks. However, the quadratic complexity of self-attention makes them computationally expensive for high-resolution document images.
Compared with CNN-based alternatives, transformer models usually provide stronger global context modeling, which is beneficial for complex page structures. A limitation of this line of work is that computational cost grows quickly with page resolution, making practical deployment more challenging.
To improve scalability, recent work explores more efficient transformer variants. For instance, additive attention mechanisms have been introduced to reduce the complexity of self-attention from quadratic to linear while maintaining strong global representation capabilities. The additive-attention-based semi-supervised framework proposed in \cite{shehzadi2025docsemi} integrates this efficient attention mechanism into a student–teacher training pipeline, enabling accurate layout detection with reduced computational overhead and limited annotated data.

\subsection{Multimodal Layout Understanding}

Although visual features provide important cues for detecting layout regions, understanding document structure often requires integrating textual and semantic information. As a result, multimodal approaches that jointly model text, layout, and visual features have become increasingly important in DLA research.

LayoutLM \cite{xu2020layoutlm} was among the first models to integrate textual content and layout information within a unified Transformer architecture. By combining word embeddings with positional and visual features, LayoutLM enables models to learn semantic relationships between document components. Subsequent models such as LayoutLMv2 and LayoutLMv3 further enhanced multimodal interactions through improved cross-modal pretraining objectives and masking strategies \cite{xu2020layoutlmv2,huang2022layoutlmv3}.
This line of work is better suited than purely visual layout detectors to scenarios in which semantics and geometry must be interpreted jointly. Such methods typically depend on reliable textual inputs, so their advantages may diminish when OCR quality is poor.

Beyond these models, recent studies explore alternative representations for encoding layout information. Grid-based approaches such as CharGrid and BERTGrid represent documents as structured grids that preserve spatial relationships between characters and tokens \cite{katti2018chargrid,denk2019bertgrid}. Graph-based models instead treat document components as nodes in a graph and model structural relationships between them \cite{wang2023graphical}. These approaches improve the ability of models to capture hierarchical and relational layout structures.

With the emergence of LLMs, document layout information is increasingly incorporated into general document understanding systems. Recent work such as LayTokenLLM \cite{zhu2025simple} proposes representing layout information as lightweight tokens that are interleaved with textual content and fed into LLMs. By introducing a single layout token per text segment and a specialized positional encoding scheme, this approach preserves textual context while enabling models to reason about document structure. Such techniques demonstrate how layout analysis can be integrated with large-scale language models to support complex tasks such as document question answering and information extraction.
Recent work has also explored leveraging layout-aware document representations for retrieval tasks. For example, ColParse~\cite{yan2026beyond} utilizes document parsing models to generate a compact set of layout-informed sub-image embeddings that are fused with global page representations, enabling efficient and structurally-aware visual document retrieval.
Rather than remaining limited to earlier task-specific architectures, these approaches broaden the role of layout modeling from detection to downstream reasoning and retrieval. They also make system behavior more dependent on cross-module alignment, which remains difficult to control in complex real-world settings.

\subsection{Emerging Trends in Layout Analysis}

Recent research has also highlighted several emerging challenges in DLA, including dataset diversity, annotation cost, and real-world document variability.

One important trend is the development of more diverse and fine-grained datasets. Early datasets such as PRImA and PubLayNet primarily contain English PDF documents with limited layout diversity. To address these limitations, M6Doc \cite{cheng2023m6doc} introduces a large-scale dataset containing multiple document formats (PDF, scanned, and photographed documents), multiple document types, and bilingual annotations. With 74 fine-grained layout categories and more than 200,000 annotations, M6Doc provides a comprehensive benchmark for fine-grained logical layout analysis.

Another active research direction is semi-supervised learning for DLA. Since manual annotation of layout datasets is expensive and time-consuming, recent methods leverage unlabeled documents to improve model performance. DocSemi \cite{shehzadi2025docsemi} proposes a DETR-based semi-supervised framework that combines one-to-one and one-to-many assignment strategies through a hybrid matching mechanism. By integrating focused attention networks and guided query strategies within a teacher–student framework, the model generates high-quality pseudo-labels and achieves competitive performance using limited labeled data.

Overall, the evolution of document layout analysis reflects a shift from rule-based page segmentation to deep learning-based visual detection and, more recently, to multimodal document understanding systems. Current research increasingly focuses on improving scalability, robustness across diverse document types, and integration with large language models. These advances are expected to play a critical role in enabling intelligent document processing systems capable of understanding complex real-world documents.

%% file: section/sec4_Optical_Character_Recognition.tex
\section{Optical Character Recognition}
\label{sec:OCR}

Optical Character Recognition (OCR) is the foundational process of converting visual text into editable digital formats. Traditionally, this field has evolved through three major technical paradigms: text detection, text recognition, and unified text spotting.

\subsection{Foundations and Classical Pipelines}
Early OCR research primarily followed a modular pipeline. Text Detection was treated as a specialized case of object detection or instance segmentation. Methods ranged from regression-based approaches like TextBoxes~\cite{liao2017textboxes} and EAST~\cite{zhou2017east}, which directly predict bounding boxes, to segmentation-based methods like PixelLink~\cite{deng2018pixellink} and CRAFT~\cite{baek2019character} that handle irregular shapes by classifying pixels or character regions.
Segmentation-based methods are usually more robust than regression-based detectors to curved or irregular text. They often require more elaborate post-processing, however, to recover coherent text instances.

Text Recognition subsequently transcribes these localized regions into character sequences. Classical models utilized Connectionist Temporal Classification (CTC) loss, exemplified by the CRNN architecture~\cite{shi2016end}, to handle sequence alignment without explicit segmentation. With the rise of attention mechanisms, Sequence-to-Sequence (Seq2Seq) models~\cite{cheng2018aon} and Transformer-based architectures like TrOCR~\cite{li2023trocr} further integrated visual features with linguistic context, significantly improving accuracy on distorted or blurred text.
Compared with CTC-based models, attention-based recognizers generally provide greater flexibility for irregular text and long-range dependencies. This advantage, however, is often accompanied by higher decoding complexity and a greater tendency to generate unstable outputs under severe noise.

To mitigate error propagation between these two isolated stages, Text Spotting emerged as an end-to-end paradigm. These frameworks, such as ABCNet~\cite{liu2020abcnet} and SwinTextSpotter~\cite{huang2022swintextspotter}, unify detection and recognition by sharing feature representations, allowing for joint optimization and more robust performance in complex scene text environments.
Text spotting improves consistency between localization and transcription relative to modular pipelines. A limitation is that joint optimization can make diagnosis and correction of failure cases less straightforward.

\subsection{The Transition to Multimodal Intelligence}
With the rapid maturation of general-purpose large vision-language models (LVLMs) and specialized end-to-end document parsing architectures, these models have demonstrated strong capabilities in detecting and recognizing text within complex scenes, historical manuscripts, and distorted documents. Their performance on OCR-centric tasks, such as visual question answering (VQA), has shifted research attention. Today, general-purpose models can often achieve highly competitive OCR performance as a byproduct of multimodal alignment, reducing the relative emphasis on standalone OCR systems.

\subsection{Modern Research Frontiers}
Despite the shift toward general models, independent OCR research remains vital in specific frontiers, focusing on extreme scenario adaptation, interpretability, and output reliability. Current trends in OCR development primarily manifest in the following directions:

\begin{itemize}
\item \textbf{Vertical and Complex Scenarios:} Recent efforts prioritize specialized domains where general VLMs may still falter~\cite{pandey2025survey}. For instance, DeepAd-OCR~\cite{lu2025deepad} leverages AI-enhanced OCR to optimize conversion elements in digital advertisements in real-time, integrating deep reinforcement learning to balance recognition accuracy with business metrics like conversion rates and regulatory compliance. Similar advancements are seen in low-resource languages, historical document restoration, and high-precision invoice parsing.
These specialized systems can better accommodate domain constraints and evaluation targets than general-purpose models. Their gains are often tied to narrower data distributions, which may limit transferability across document types.

\item \textbf{Internal Mechanism and Interpretability:} As OCR capabilities are absorbed by large models, understanding how these models "read" becomes crucial. Research into OCR Heads~\cite{baek2025large} identifies specialized attention units within LVLMs that are distinct from standard text-retrieval heads. These OCR-specific heads focus on visual patches to guide text extraction, offering a mechanistic path to reduce hallucinations and improve grounding in multimodal reasoning.
This line of work differs from performance-oriented OCR research by focusing on model behavior rather than only benchmark gains. Its current limitation is that mechanistic findings do not yet translate directly into standardized improvements across diverse OCR settings.

\item \textbf{Uncertainty Quantification and Secondary Optimization:} Given the "black-box" nature of large models, ensuring the reliability of OCR outputs is a major challenge. The Consensus Entropy (CE)~\cite{zhang2025consensus} framework introduces a novel uncertainty metric based on inter-model agreement. By calculating the semantic divergence among multiple VLM predictions, CE enables adaptive routing—merging high-confidence results while redirecting high-entropy (ambiguous) cases to more powerful specialized models for secondary verification.
Relative to single-pass inference, this strategy provides a more practical way to trade off cost and reliability. It also introduces additional system complexity and depends on the availability of well-calibrated fallback models.

\end{itemize}

In summary, while OCR is no longer viewed solely as a simple ``detection-then-recognition'' task, the field is evolving toward more interpretable, reliable, and scene-adaptive text perception within the broader document AI ecosystem.

%% file: section/sec5_Mathematical_Formula_Detection_and_Recognition.tex
\section{Mathematical Expression Detection and Recognition} 
\label{sec:MER}

Mathematical expressions are essential components of scientific and technical documents, widely appearing in domains such as mathematics, physics, and engineering. Compared with ordinary text, mathematical expressions pose unique challenges for document understanding systems due to their large symbol vocabulary, two-dimensional spatial layouts, and complex hierarchical structures. Accurately extracting mathematical expressions from documents is therefore a critical step toward the digitization and semantic understanding of scientific knowledge.

\begin{figure}[t]
\centering
\includegraphics[width=1.0\textwidth]{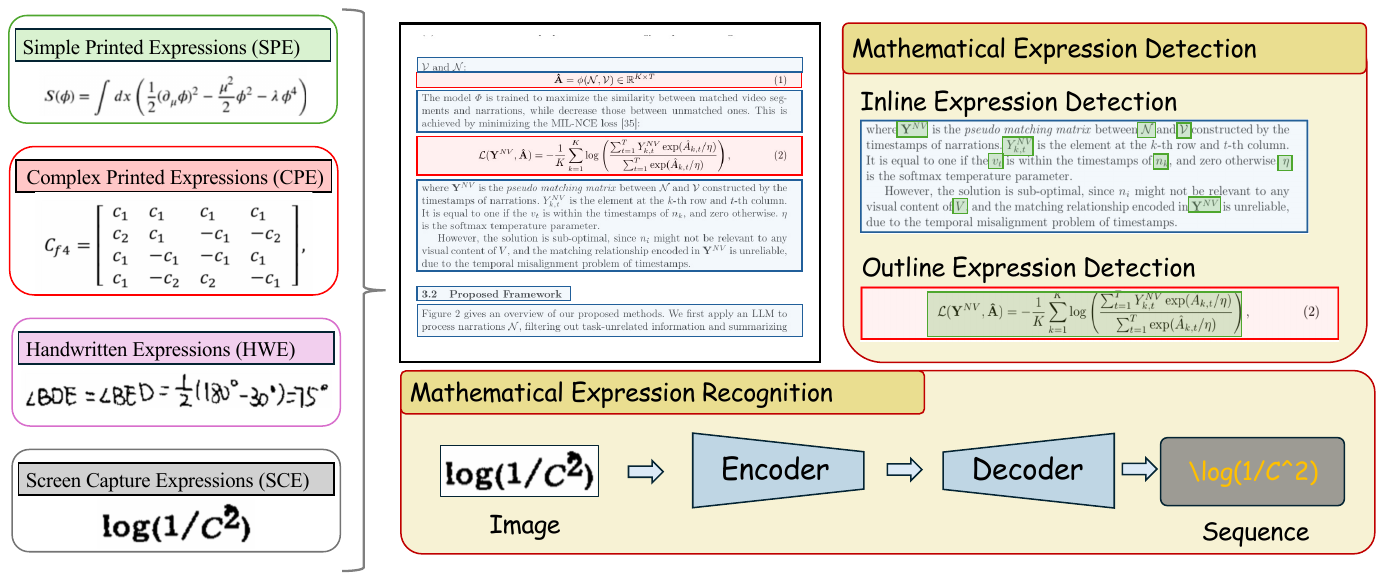}
\vspace{-5mm}
\caption{Overview of the mathematical expression detection and recognition pipeline. The figure distinguishes region localization from downstream conversion into structured markup, highlighting the progression from expression detection to structure-aware recognition.} 
\vspace{-6mm}
\label{fig5}
\end{figure}

Figure~\ref{fig5} illustrates a typical pipeline for mathematical expression detection and recognition. The processing of mathematical expressions generally involves two stages: \textit{detection} and \textit{recognition}. Detection aims to locate mathematical expression regions within document images, while recognition converts the detected mathematical expression images into structured markup representations such as \LaTeX{}. Mathematical expressions typically appear in two forms: \textit{displayed mathematical expressions}, which are visually separated from surrounding text, and \textit{inline mathematical expressions}, which are embedded within text lines and therefore harder to identify.

Research on mathematical expression recognition dates back to the 1960s \cite{anderson1967syntax}. Over the past decades, the field has evolved from rule-based parsing methods to modern deep neural models. 

\subsection{Mathematical Expression Detection}

Early mathematical expression detection (MED) methods treated mathematical expressions as visual objects and employed CNN or handcrafted features to locate expression regions \cite{gao2017deep, yi2017cnn, li2018page}. Later studies adapted general object detection frameworks such as SSD and YOLO to detect mathematical expressions in document images \cite{mali2020scanssd, nguyen2021ds}. Two-stage detectors, including Faster R-CNN and Mask R-CNN, were also explored for higher localization accuracy \cite{younas2019ffd}. More recent work incorporates contextual and structural cues, such as FormulaDet \cite{hu2024mathematical}, which formulates mathematical expression detection as an entity–relation extraction problem and leverages layout-aware modeling to improve robustness.
Context-aware methods are better aligned than generic object detectors with the ambiguity of inline and densely arranged formulas. This added structural modeling also increases task complexity and annotation requirements.

However, with the development of unified document understanding frameworks, mathematical expression detection is rarely treated as an independent task. Modern document parsing systems typically detect mathematical expressions together with other layout elements (e.g., text blocks, tables, and figures) through document layout analysis models. As a result, recent research increasingly focuses on improving mathematical expression recognition and structural parsing rather than standalone detection algorithms.

\subsection{Mathematical Expression Recognition}

Mathematical Expression Recognition (MER) aims to convert an image of a mathematical expression into a machine-readable representation such as LaTeX. Compared with standard OCR, MER must not only recognize individual symbols but also infer their spatial and hierarchical relationships, making the task significantly more challenging.

Early systems relied on handcrafted grammars and rule-based parsing strategies. With the advancement of deep learning, MER has largely shifted to neural encoder–decoder frameworks that treat the problem as an image-to-sequence generation task. Recent work further enhances these models by incorporating structural reasoning, improved visual representations, and multimodal learning techniques.

\subsubsection{Encoder–Decoder Models and Sequence-based Recognition}

Most modern MER systems adopt encoder–decoder architectures that transform visual features into symbolic sequences. Early deep learning models introduced attention-based sequence decoding for handwritten mathematical expressions, significantly improving recognition performance.
Subsequent studies focused on improving visual feature extraction and sequence modeling. Dense convolutional networks and multi-scale encoders were widely adopted to better capture symbol appearance and spatial context. Transformer-based decoders were later introduced to model long-range dependencies more effectively. For instance, BTTR \cite{zhao2021handwritten} employs a Transformer decoder for handwritten mathematical expression recognition, while CoMER \cite{zhao2022comer} improves symbol alignment and decoding accuracy through refined attention mechanisms.
Relative to earlier recurrent decoders, these models improve long-range dependency modeling and are better suited to complex handwritten expressions. They still linearize inherently two-dimensional structures, however, which limits their ability to represent hierarchical relations explicitly.
Despite their success, purely sequence-based models often struggle to fully capture the hierarchical structure of mathematical expressions, since the underlying representation is inherently two-dimensional while the output representation is linearized.

\subsubsection{Structure-aware Modeling}

To address the structural complexity of mathematical expressions, recent work incorporates explicit structural modeling into the recognition process.
One line of research models expressions using hierarchical structures. For example, TAMER \cite{zhu2025tamer} introduces a tree-aware Transformer architecture that jointly learns sequence prediction and expression tree structures. The proposed Tree-Aware Module enhances the model’s ability to capture hierarchical relationships while maintaining efficient training.
Compared with purely sequence-based recognition, this direction offers a more natural inductive bias for mathematical syntax. A limitation is that gains in structural fidelity may come with increased modeling complexity and more specialized supervision.
Another direction focuses on modeling spatial relationships between symbols. SSAN \cite{zhang2025ssan} introduces a symbol spatial-aware network that predicts spatial distribution maps for symbols as an auxiliary task, allowing the model to better capture two-dimensional layouts. Other methods integrate structural constraints directly into the decoding process. The Structure and Counting Aware Network (SCAN) \cite{li2022counting} proposes a Skeleton Shaping and Character Counting Module that simultaneously predicts expression skeleton structures and symbol frequency distributions, enabling the model to correct potential symbol misrecognition caused by visually similar characters.
These approaches are more effective than generic encoder--decoder models when symbol arrangement carries substantial semantic meaning. They may be less flexible, however, when expression styles vary widely across domains or writing conditions.
Some studies also address symbol ambiguity through explicit semantic supervision. The Type-Aware Network (TAN) \cite{xie2025tan} incorporates symbol-type labels into an attention-based encoder–decoder framework. By introducing explicit supervision on symbol categories, the model improves symbol-level feature representation and recognition robustness for visually similar handwritten characters.

\subsubsection{Visual Representation and Scaling}

Beyond decoder design, recent studies highlight the importance of robust visual representations. Mathematical expressions often exhibit large variations in scale and aspect ratio, which can lead to information loss when using fixed-resolution inputs.
AutoScaler \cite{yang2025autoscaler} identifies this scale misalignment problem in mathematical expression recognition and proposes a self-scale alignment framework that adaptively determines the optimal input scale for each mathematical expression image. By introducing stochastic scale training and scale-aware inference, the model improves recognition robustness across expressions with diverse structural complexity.
Compared with fixed-resolution pipelines, adaptive scaling is better suited to expressions with large variation in length and symbol density. Its practical limitation is that dynamic scaling may complicate inference control and latency optimization.

\subsubsection{Data and Efficiency Considerations}

The progress of mathematical expression recognition has also been constrained by the limited size of existing datasets. Traditional benchmarks such as CROHME contain relatively small numbers of training samples, which restrict model generalization.
To address this limitation, the MathWriting dataset \cite{gervais2025mathwriting} introduces more than 600k handwritten mathematical expressions, providing one of the largest publicly available datasets for mathematical expression recognition research. Large-scale datasets enable the training of more powerful models and improve generalization across diverse handwriting styles.
In addition to dataset expansion, recent work also focuses on improving computational efficiency for real-world applications. PP-FormulaNet \cite{liu2025pp} constructs a large-scale mathematical expression corpus from arXiv papers and proposes a lightweight recognition framework combining knowledge distillation and multi-token prediction, achieving both higher accuracy and faster inference. Similarly, UniRec-0.1B \cite{du2025unirec} proposes a compact unified recognition model capable of handling both text and mathematical expressions, facilitating efficient document parsing systems.
For deployment-oriented settings, these compact models are often more suitable than larger specialized recognizers. Model compression and unification may nevertheless reduce headroom on highly complex expressions that require finer structural reasoning.

\subsubsection{Vision-Language Models for Mathematical Expression Recognition}

With the rapid development of multimodal large models, recent research explores applying VLMs to mathematical expression recognition. These approaches leverage large pretrained multimodal models to improve generalization and reasoning capabilities.
Uni-MuMER \cite{li2025uni} proposes a unified multi-task fine-tuning framework that enhances open-source VLMs for handwritten mathematical expression recognition. By integrating auxiliary tasks such as symbol counting and tree-aware reasoning, the model improves structural understanding and achieves strong performance across multiple datasets.
DocTron-Formula \cite{zhong2025doctron} further demonstrates that multimodal large models can handle complex real-world mathematical expressions after supervised fine-tuning on dedicated datasets containing multi-line expressions.
HiE-VL \cite{guo2025hie} represents another attempt to adapt large vision-language models for handwritten mathematical expression recognition. It introduces a hierarchical adapter architecture consisting of primitive-level and structural adapters to better capture fine-grained visual features and hierarchical relationships within mathematical expression images, together with a progressive training strategy that gradually improves visual perception and mathematical language understanding.
VLM-based methods offer broader transferability and stronger multimodal priors than conventional MER architectures. They still require careful adaptation, however, to control output stability and avoid structurally plausible but incorrect generations.

\medskip

Overall, research on mathematical expression recognition has evolved from rule-based parsing to neural encoder–decoder models, and more recently to structure-aware architectures and multimodal large-model approaches. Current studies mainly focus on challenging scenarios such as handwritten mathematical expressions and long or structurally complex expressions. Correspondingly, many methods attempt to improve recognition accuracy by incorporating explicit structural modeling, spatial reasoning, symbol-level supervision, and improved visual representations. Meanwhile, the construction of large-scale and more challenging datasets has become an important driver of progress in this field. In addition, some recent works explore refinement or post-processing strategies that iteratively correct initial predictions, further improving the robustness and accuracy of mathematical expression parsing in real-world document understanding systems.

%% file: section/sec6_Table_Detection_and_Recognition.tex
\section{Table Detection and Recognition} 
\label{sec:table}

\begin{figure}
\centering
\includegraphics[width=1.0\textwidth]{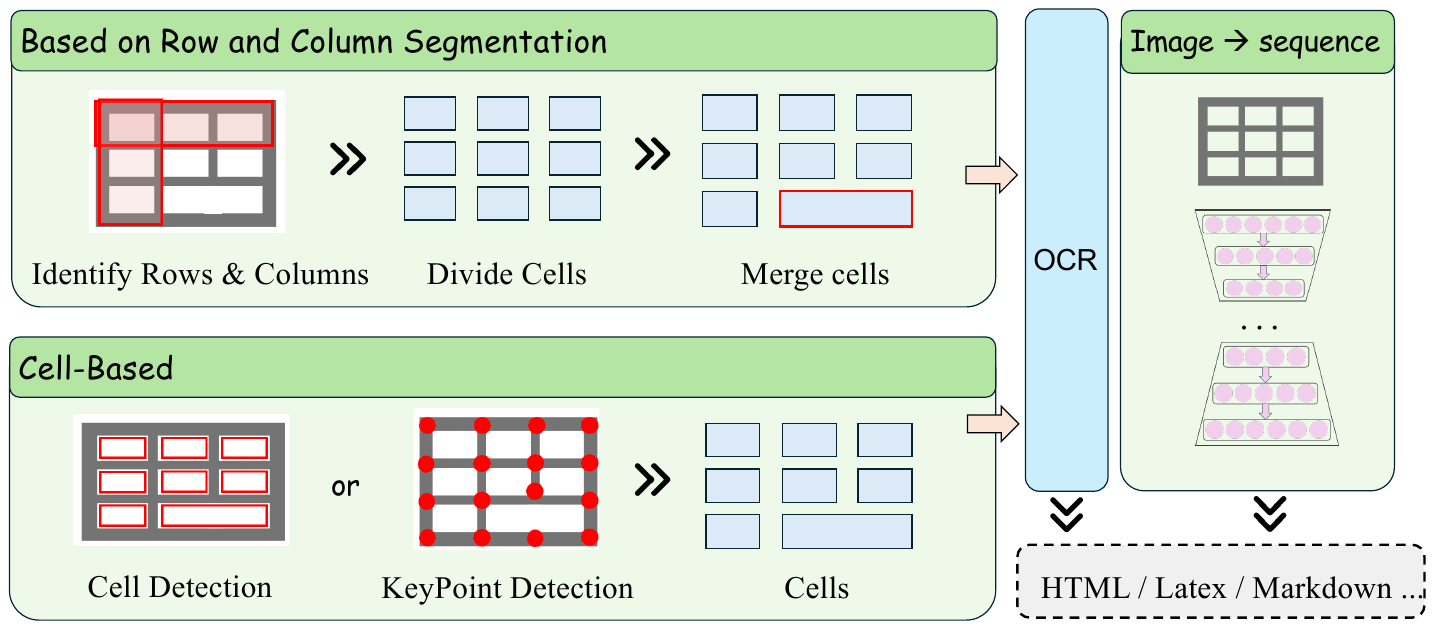}
\caption{Overview of table detection and recognition. The figure summarizes the transition from table region localization to structure parsing and content extraction, emphasizing the dependencies among detection, cell organization, and structured output generation.} 
\label{fig6}
\end{figure}

Tables provide structured data representations that facilitate rapid understanding of relationships and hierarchies. Accurate table detection and recognition are therefore crucial for effective document analysis, especially in specific scenarios in academic and financial domains~\cite{zhang2024ocr, yang2025cc}. Figure~\ref{fig6} presents an overview of the main algorithmic pipeline.

Table detection involves identifying and segmenting table areas within document images or electronic files. The goal is to locate tables and distinguish them from other content, such as text or images.

With improvements in detection accuracy, research has shifted toward table structure recognition. This task involves analyzing the internal structure of tables after detection, including segmenting rows and columns, extracting cell content, and interpreting cell relationships in structured formats such as LaTeX.

This section reviews object-detection-based algorithms for table detection and discusses representative deep-learning-based table recognition methods from recent research. 

\subsection{Table Detection Based on Object Detection Algorithms} 

Table detection (TD) is often approached as an object detection task, where tables are treated as objects, using models originally designed for natural images. Despite differences between page elements and natural images, one-stage, two-stage, and transformer-based models can achieve robust results with careful retraining and tuning, often serving as benchmarks for TD. In practice, due to the rapid progress of Document Layout Analysis (DLA), table detection has achieved relatively mature performance on many benchmarks, and research efforts have increasingly shifted toward more challenging tasks such as table structure understanding and table recognition.

To adapt object detection for TD, various studies have enhanced standard methods. For instance, \cite{hao2016table} integrates PDF features, like character coordinates, into CNN-based models. \cite{gilani2017table} customizes Faster R-CNN for document images by modifying representation and optimizing anchor points. \cite{schreiber2017deepdesrt} combines Deformable CNNs with Faster R-CNN to handle varying table scales, while \cite{siddiqui2018decnt} fine-tunes Faster R-CNN specifically for tables. \cite{huang2019yolo} employs the YOLO series, enhancing anchor and post-processing techniques.
These adaptations exploit document-specific cues such as text alignment and table geometry more effectively than generic detectors. They still inherit the limitations of object detection formulations, especially when borders are faint or table semantics are ambiguous.

To address table sparsity, \cite{xiao2023table} expands SparseR-CNN with Gaussian Noise Augmented Image Size proposals and many-to-one label assignments, introducing the Information Coverage Score (ICS) to evaluate recognition accuracy.
This line of work is particularly useful for sparse or weakly bounded tables, where standard detectors tend to miss large empty regions. A limitation is that better region coverage does not by itself resolve downstream structure parsing errors.

\subsection{Table Recognition}

Traditionally, table structure recognition relied on manually designed rules and heuristics, such as Hough Transform-based line detection or whitespace analysis for borderless tables. These rule-based methods were often limited when dealing with complex layouts or irregular table formats. With the development of machine learning and deep learning, table recognition gradually evolved toward data-driven approaches capable of modeling richer spatial relationships.

Early deep learning studies often decomposed the problem into multiple subtasks, including row and column detection, cell segmentation, and OCR-based text extraction. These modular pipelines improved robustness compared with rule-based systems but also introduced challenges such as error propagation between stages. More recent research has explored unified and end-to-end solutions that jointly model table structure and content.

TabNet~\cite{arik2021tabnet} is a pioneering deep learning model for table feature extraction, handling both numerical and categorical features in an end-to-end fashion. It features an efficient and interpretable learning architecture, optimized for various tasks. TabNet’s sequential attention mechanism allows the model to focus on relevant features progressively, using instance-level sparse feature selection and a multi-step decision process. This enhances TabNet's ability to explain feature importance at both local and global levels. Building on this, models like TabTransformer~\cite{huang2020tabtransformer} have further advanced table feature extraction, providing valuable insights for developing robust table recognition models.

\subsubsection{Methods Based on Row and Column Segmentation}

A key challenge in table structure recognition is detecting individual cells, particularly in the presence of large blank spaces. Early deep learning approaches addressed this by segmenting tables into rows and columns. These algorithms generally adopt a top-down strategy, first identifying the overall table region and then segmenting it into rows and columns. This method is effective for tables with clear boundaries and simple layouts.

\begin{itemize}
\item \textbf{Row and Column Detection:}
Initially, table structure recognition was seen as an extension of table detection, primarily using object detection algorithms to identify table bounding boxes. Segmentation algorithms then established relationships between rows and columns. CNN and transformer architectures were pivotal in this context~\cite{siddiqui2019deeptabstr, zou2020deep}.

Transformers, such as DETR, excel at recognizing global relationships within an image, enhancing generalization. Innovations include row and column segmentation through transformer queries~\cite{guo2022trust} and a dynamic query enhancement model, DQ-DETR~\cite{wang2023robust}. Additionally, Bi-directional Gated Recurrent Units (Bi-GRUs) effectively captured row and column separators by scanning images bidirectionally~\cite{khan2019table}.
Relative to purely CNN-based segmentation, these methods generally offer stronger modeling of global row--column interactions. They remain most effective, however, on tables whose structural regularity can still be decomposed into explicit separators.

\item \textbf{Fusion Module:}
Earlier methods focused on detecting table lines but often overlooked complex inter-cell relationships. Advanced algorithms now estimate merging probabilities between cells to improve recognition accuracy in tables without explicit row and column lines. For example, embedding modules integrate plain text within grid contexts to guide merge predictions via GRU decoders~\cite{zhang2022split}. Other techniques use adjacency criteria and spatial compatibility to predict cell mergers~\cite{lin2022tsrformer}. The integration of global computational models, such as Transformers, further enhances the analysis of complex tables~\cite{nguyen2023formerge}.
Compared with line-centric methods, fusion-based approaches are better suited to borderless tables and merged cells. Their limitation is that performance becomes more sensitive to the quality of intermediate text and cell representations.

\end{itemize}

CNNs remain foundational for feature extraction in table images, although recent efforts aim to optimize architectures for table-specific characteristics. For example, replacing ResNet18 with ShuffleNetv2 significantly reduced model parameters~\cite{zhang2023table}. Despite progress, challenges persist in tables that lack explicit lines, such as those with sparse content or irregular arrangements.

\subsubsection{Methods Based on Cells}

Cell-based methods, often described as bottom-up approaches, construct table structures by first detecting individual cells and then modeling the relationships between them. These approaches usually involve detecting cell boundaries followed by structural reconstruction, which is particularly useful for tables with irregular layouts.
Early improvements mainly focused on enhancing cell detection accuracy. For instance, HRNet-based backbones have been applied for high-resolution feature extraction in table segmentation tasks~\cite{prasad2020cascadetabnet}. Some studies introduced specialized loss functions to improve detection quality, including continuity and overlap constraints~\cite{raja2022visual}. Other works proposed dual-path architectures to jointly learn local features and segmentation masks~\cite{nguyen2022tablesegnet}.
Another direction focuses on geometric representations of cells. Vertex prediction methods detect the corners of cells to better handle perspective distortions or irregular table boundaries. For example, the Cycle-Pairing Module predicts both centers and vertices of cells simultaneously~\cite{long2021parsing}. Graph-based approaches further represent tables as graphs, where cells are nodes and spatial relationships define edges, enabling Graph Neural Networks (GNNs) to model complex adjacency relationships~\cite{chi2019complicated, qasim2019rethinking}.
Bottom-up cell modeling is often more flexible for irregular layouts than top-down row--column segmentation. This flexibility comes at the cost of stronger dependence on accurate low-level cell localization.
Although these methods can effectively capture structural relationships, their performance often depends heavily on accurate cell detection, and errors in early stages may propagate to later structural inference.

\subsubsection{Image-to-Sequence Approaches}

Inspired by advances in OCR and formula recognition, image-to-sequence methods reformulate table recognition as a sequence generation task, converting table images directly into structured representations such as LaTeX or HTML. Encoder–decoder architectures with attention mechanisms encode visual features and generate markup sequences describing table structure and content.
Early studies such as \cite{deng2019challenges} explored encoder–decoder models for converting scientific table images into LaTeX code. Later works extended these models with dual-decoder architectures to handle structural tags and textual content simultaneously~\cite{zhong2020image}. Architectures such as MASTER~\cite{ye2021pingan} further improved sequence modeling capabilities for complex table layouts.
Image-to-sequence approaches simplify the interface between structure analysis and content generation relative to modular reconstruction pipelines. They can still be sensitive to output linearization choices and may struggle when multiple structural interpretations are plausible.
More recent research has explored transformer-based architectures for table parsing. For example, TransTab~\cite{wang2025transtab} introduces Vision Transformers (ViT) to model long-range dependencies within table layouts, improving the detection of complex row and column relationships. By combining dedicated modules for table detection and column localization with OCR-based text extraction, the model achieves improved robustness on complex tables containing multi-row or multi-column spans.
A key challenge in traditional pipelines is the alignment between detected table structures and OCR-extracted text. Small errors in cell bounding boxes may lead to incorrect pairing between structure tokens and text content. To address this issue, recent work proposes end-to-end recognition frameworks that directly generate both structural tokens and cell content. For example, E2eTRNet~\cite{yang2025alignment} introduces a dual-decoder architecture where semantic features associated with structure tokens guide the prediction of cell content, enabling automatic alignment and eliminating the need for explicit bounding-box matching.
These alignment-aware frameworks improve consistency between structure and content relative to loosely coupled OCR-plus-structure pipelines. They also concentrate multiple sources of difficulty within a single model, which can make optimization less stable.
Another emerging trend is to unify table recognition under a language modeling framework. UniTable~\cite{peng2024unitable} reformulates table recognition as an image-to-text task, jointly predicting table structure, cell content, and bounding boxes using a unified sequence generation objective. The framework leverages self-supervised pretraining on large-scale unlabeled table images, significantly improving generalization performance across multiple benchmarks.
More recently, VLMs have shown strong capabilities for document understanding tasks, including table recognition. Instead of relying solely on labeled datasets, TRivia~\cite{zhang2025trivia} proposes a self-supervised fine-tuning framework that enables VLMs to learn table recognition from large collections of unlabeled table images. The method introduces a question-answering-based proxy task and reinforcement learning optimization to generate supervisory signals automatically, allowing models to improve recognition performance without additional human annotations.
Compared with task-specific recognizers, VLM-based approaches offer broader transferability and weaker reliance on manual annotation. A limitation of this paradigm is that structural faithfulness may still lag behind specialized models on complex spanning and long tables.
Overall, table recognition research has gradually evolved from rule-based methods and modular detection pipelines toward unified, end-to-end frameworks that integrate visual perception, language modeling, and large-scale self-supervised learning. These developments have significantly improved robustness in handling complex table layouts and real-world document scenarios.

%% file: section/sec7_special_elements_parsing.tex
\section{Visual Element Parsing}
\label{sec:other_elements}

Beyond standard textual, formulaic, and tabular data, scientific and technical documents often contain specialized visual elements that encode high-density domain knowledge. Extracting structured information from these graphical components is essential for comprehensive document understanding and knowledge graph construction. This section focuses on two critical categories of non-textual elements: charts (Sec.~\ref{sec:chart}), which represent quantitative data trends, and chemical structures (Sec.~\ref{sec:OCSR}), which depict the spatial and symbolic arrangements of molecular entities. We discuss the evolution of parsing techniques for these elements, moving from traditional rule-based pipelines to modern multimodal generative frameworks.

\subsection{Chart Parsing}
\label{sec:chart}

Charts are widely used in scientific publications, business reports, and technical documents to visually summarize structured data and reveal patterns, trends, and comparisons~\cite{davila2020chart}. Unlike plain text or tables, charts encode information through graphical marks such as bars, lines, symbols, colors, and spatial layouts. Chart parsing aims to automatically convert these visual representations into machine-readable structured formats. By generating structured outputs, such as tables (e.g., Markdown or HTML), JSON schemas, or visualization description languages such as \texttt{TikZ}, chart parsing enables the extraction of explicit semantic information from visualizations~\cite{rong2025review,xiao2025adaptive}. These representations provide structured context for downstream tasks such as chart reasoning, document understanding, and multimodal information retrieval, making chart parsing an important component of document intelligence systems.

Early research on chart understanding mainly relied on pipeline-based computer vision systems that detect chart elements and reconstruct the underlying data structure. These methods typically involve several stages, including chart classification~\cite{dhote2024swin,thiyam2024chart,shaheen2024c2f}, element detection~\cite{poco2017reverse,xu2024empowering}, text recognition through OCR~\cite{qiao2023structure}, and heuristic rule-based reasoning to associate visual elements with textual labels~\cite{davila2022icpr,mustafa2023charteye}. Although such approaches achieved reasonable performance for simple charts, they often suffered from error propagation and limited robustness across diverse chart styles and layouts. 
These pipelines are usually more interpretable and easier to debug stage by stage than recent generative frameworks. Their modular nature nevertheless makes them vulnerable to accumulated errors when chart elements are densely coupled.

Recent advances in vision-language models have shifted chart parsing toward end-to-end multimodal generation frameworks, where models directly generate structured representations from chart images instead of relying on explicit element detection. A central challenge in this paradigm is aligning chart images with structured data formats while preserving both numerical and visual semantics. ChartAssistant~\cite{meng2024chartassistant} addresses this issue by introducing a chart-to-table pre-training task that converts charts into Markdown tables, enabling the model to learn structural relationships between graphical elements and numerical values before multitask instruction tuning. Beyond structural alignment, efficiency and quantitative reasoning also become critical challenges. TinyChart~\cite{zhang2024tinychart} improves chart understanding by reducing redundant visual tokens through a visual token merging strategy and enhancing numerical reasoning via Program-of-Thoughts learning, where the model generates executable Python programs to perform intermediate calculations. More recently, ChartMoE~\cite{xu2024chartmoe} explores richer structured representations by aligning charts with multiple modalities, including tables, JSON attributes, and visualization code, using a mixture-of-experts connector initialized with diverse alignment tasks. This design allows the model to capture complementary chart semantics such as numerical data, layout attributes, and rendering information.
Relative to earlier pipelines, these methods improve end-to-end flexibility and reduce the need for manually designed association rules. Direct generation also makes numerical faithfulness harder to guarantee, especially for visually cluttered or stylistically unusual charts.

In addition to structured data extraction, an emerging direction in chart parsing is \textit{chart-to-code} generation, where charts are converted into executable visualization programs. Compared with tables or JSON representations, plotting code provides a nearly lossless description of charts because it encodes both data and rendering logic. VinciCoder~\cite{zhao2025vincicoder} explores multimodal code generation models that directly produce visualization code from chart images and improves visual fidelity through a visual reinforcement learning framework that optimizes generated programs based on the similarity between rendered and target charts. To better evaluate such capabilities, the Chart2Code~\cite{tang2025charts} benchmark introduces a hierarchical evaluation framework with three progressively challenging tasks—chart reproduction, chart editing, and long-table-to-chart generation—revealing that current multimodal models perform well on simple reproduction but still struggle with complex editing and long-context data visualization scenarios.

In summary, chart parsing research has evolved from traditional rule-based pipelines toward end-to-end multimodal generation frameworks, paving the way for more capable document parsing systems.

\subsection{Optical Chemical Structure Recognition}
\label{sec:OCSR}

Optical Chemical Structure Recognition (OCSR) is a specialized task within document parsing that aims to convert graphical depictions of chemical molecules into machine-readable formats, such as SMILES, InChI, or connection tables~\cite{musazade2022review}. In scientific literature and patents, chemical structures are not merely decorative but serve as the primary medium for conveying molecular topology, stereochemistry, and reactive sites. Unlike general OCR, OCSR must parse a complex interplay of alphanumeric characters (atomic symbols), geometric primitives (bonds), and abstract notations (wedges, dashes, and aromatic rings). 
% Accurately parsing these structures is a prerequisite for automated drug discovery, patent mining, and the digitization of chemical knowledge.

Early research on OCSR primarily followed a graph reconstruction paradigm, utilizing hand-crafted image processing algorithms to identify atoms and bonds as nodes and edges. Tools such as OSRA \cite{filippov2009optical} and chemoCR~\cite{zimmermann2011chemical} relied on binarization, skeletonization, and heuristic rules to assemble the molecular graph. While these methods provided a foundation for chemical informatics, they were highly sensitive to image artifacts, such as broken lines or overlapping text, which are common in scanned historical documents~\cite{musazade2022review}.

With the rise of deep learning, OCSR has been reformulated as a multimodal sequence generation task, analogous to image captioning. Recent advances have seen the adoption of Transformer-based architectures to overcome the limitations of traditional pipelines. SwinOCSR~\cite{xu2022swinocsr} employs a Swin Transformer as a hierarchical vision backbone to capture multi-scale structural features, converting them into DeepSMILES strings. To further improve feature representation, MPOCSR~\cite{lin2024mpocsr} introduces a multi-path Vision Transformer (MPViT) and a class-balanced loss function to mitigate the long-tail distribution of chemical elements. For scenarios with limited supervision, such as hand-drawn sketches, recent work~\cite{oldenhof2024atom} explores atom-level entity localization combined with self-relabeling strategies, demonstrating that explicit spatial grounding can significantly improve data efficiency compared to pure string-based generation.

The latest frontier in OCSR involves the integration of Multimodal Large Language Models (MLLMs) and specialized semantic optimization. As molecular complexity increases, standard SMILES often fails to represent specialized entities like Markush structures in patents. MolParser~\cite{fang2025molparser} addresses this by introducing an extended SMILES (E-SMILES) format and a large-scale dataset (MolParser-7M), enabling robust parsing of varied drawing styles in the wild. Meanwhile, models like ChemVLM~\cite{li2025chemvlm} and TinyChemVL~\cite{zhao2025tinychemvl} expand the scope of OCSR from recognition to chemical reasoning, with the latter utilizing visual token pruning to achieve high-speed inference (up to 40 FPS). To ensure the chemical validity of generated outputs, MolSight~\cite{zhang2025molsight} introduces a reinforcement learning framework using the GRPO algorithm, which optimizes the model based on chemical semantic correctness rather than just token-level accuracy. This allows for superior recognition of challenging stereoisomers that share identical connectivity but different spatial configurations.
Unlike earlier sequence-generation models, these approaches place greater emphasis on semantic validity rather than only syntactic matching. A remaining limitation is that stronger reasoning capability does not fully eliminate sensitivity to rare notation conventions and low-quality scans.

In summary, OCSR has progressed from fragile rule-based systems to robust, semantic-aware multimodal frameworks, enabling the seamless integration of chemical visual data into the modern digital research ecosystem.

%% file: section/sec8_Large_Language_Model_for_Document_Content_Extraction.tex
\section{VLMs for Document Parsing}
\label{sec:VLM}

VLMs have rapidly reshaped the landscape of document parsing by unifying visual perception and language modeling within large-scale multimodal architectures. This section reviews three major paradigms that have emerged in this evolution: (1) general-purpose VLMs adapted to document understanding, (2) end-to-end specialized document VLMs that directly generate structured representations, and (3) multi-stage or hybrid architectures that reintroduce controlled decomposition to improve efficiency and robustness. Together, these paradigms illustrate a broader transition from generic multimodal reasoning to structure-aware and task-aligned document intelligence.
Table~\ref{tab:omnidocbench_extended} further presents recent comprehensive VLMs for document parsing on OmniDocBench-v1.5~\cite{ouyang2024omnidocbench,duan2026glm,wu2026firered}, showing that specialized end-to-end and multi-stage document VLMs generally outperform general-purpose VLMs on structure-sensitive metrics such as formula, table, and reading-order evaluation.

Beyond the three mainstream paradigms, recent work explores hallucination mitigation~\cite{he2025seeing}, reasoning--tool integration~\cite{chen2025dianjin}, structured post-OCR correction~\cite{shim2025revise}, and fine-grained quality assessment~\cite{zhang2025docr, tan2025ocr}, collectively extending document parsing toward more reliable and controllable real-world deployment.

\begin{table*}[t]
\centering
\caption{Detailed Performance of VLMs for Document Parsing on OmniDocBench-v1.5.}
\label{tab:omnidocbench_extended}
\resizebox{\textwidth}{!}{% 
    \begin{tabular}{lcccccccc} % 增加到 9 列
    \toprule
    \textbf{Model} & \textbf{Data} & \textbf{Param} & \textbf{Overall} $\uparrow$ & \textbf{Text$^{\text{Edit}}$} $\downarrow$ & \textbf{Formula$^{\text{CDM}}$} $\uparrow$ & \textbf{Table$^{\text{TEDS}}$} $\uparrow$ & \textbf{Table$^{\text{TEDS\_s}}$} $\uparrow$ & \textbf{R-order$^{\text{Edit}}$} $\downarrow$ \\ \midrule
    
    % 第一类行
    \rowcolor[HTML]{EFEFEF} \multicolumn{9}{c}{\textit{General-Purpose VLMs}} \\ \midrule
    GPT-4o~\cite{hurst2024gpt4o} \faLock & 2024.5 & - & 75.02 & 0.217 & 79.70 & 67.07 & 76.09 & 0.148 \\
    GPT-5.2~\cite{singh2025gpt5} \faLock & 2025.12 & - & 85.50 & 0.123 & 86.11 & 82.66 & 87.35 & 0.099 \\
    Gemini-2.5 Pro~\cite{comanici2025gemini} \faLock & 2025.3 & - & 88.03 & 0.075 & 85.82 & 85.71 & 90.29 & 0.097 \\
    Gemini-3.0 Pro~\cite{google2025gemini3pro} \faLock & 2025.11 & - & 90.33 & 0.065 & 89.18 & 88.28 & 90.29 & 0.071 \\
    InternVL3-76B~\cite{zhu2025internvl3} & 2025.4 & 76B & 80.33 & 0.131 & 83.42 & 70.64 & 77.74 & 0.113 \\
    Qwen2.5-VL-72B~\cite{qwen2.5-VL} & 2025.2 & 72B & 87.02 & 0.094 & 88.27 & 82.15 & 86.22 & 0.102 \\
    InternVL3.5-241B~\cite{wang2025internvl35} & 2025.8 & 241B & 82.67 & 0.142 & 87.23 & 75.00 & 81.28 & 0.125 \\
    Qwen3-VL~\cite{qwen3technicalreport} & 2025.11 & 2B & 81.87 & 0.100 & 85.87 & 69.77 & 74.37 & 0.115 \\
    Qwen3-VL~\cite{qwen3technicalreport} & 2025.11 & 235B & 89.15 & 0.069 & 88.14 & 86.21 & 90.55 & 0.068 \\
    Qwen3.5~\cite{Qwen35}& 2026.3 & 397B & 90.80 & - & - & - & - & - \\ 
    
    \midrule
    \rowcolor[HTML]{EFEFEF} \multicolumn{9}{c}{\textit{End-to-End Specialized VLMs}} \\ \midrule
    Mistral OCR~\cite{mistral-ocr} & 2025.3 & - & 74.82 & 0.193 & 68.03 & 75.75 & 80.23 & 0.202 \\
    OCRFlux-3B~\cite{OCRFlux} & 2025.5 & 3B & 74.82 & 0.193 & 68.03 & 75.75 & 80.23 & 0.202 \\
    POINTS-Reader~\cite{liu2025points} & 2025.9 & 3B & 80.98 & 0.134 & 79.20 & 77.13 & 81.66 & 0.145 \\
    olmOCR-7B~\cite{poznanski2025olmocr} & 2025.10 & 7B & 81.79 & 0.096 & 86.04 & 68.92 & 74.77 & 0.121 \\
    MinerU2-VLM~\cite{mineru2-2025} & 2025.5 & 0.9B & 85.56 & 0.078 & 80.95 & 83.54 & 87.66 & 0.086 \\
    Nanonets-OCR-s~\cite{Nanonets-OCR-S} & 2025.10 & 3B & 85.59 & 0.093 & 85.90 & 80.14 & 85.57 & 0.108 \\
    dots.ocr~\cite{li2025dots} & 2025.7 & 2.7B & 88.41 & 0.048 & 83.22 & 86.78 & 90.62 & 0.053 \\
    DeepSeek-OCR~\cite{wei2025deepseek} & 2025.9 & 3B & 87.36 & 0.073 & 84.14 & 85.25 & 89.01 & 0.085 \\
    % HunyuanOCR~\cite{team2025hunyuanocr} & 2025.12 & 1B & 94.10 & 0.073 & 84.14 & 85.25 & 89.01 & 0.085 \\ 0.042 94.73 91.81
    DeepSeek-OCR 2~\cite{wei2026deepseek2} & 2026.2 & 3B & 91.09 & 0.048 & 90.31 & 87.75 & 92.06 & 0.057 \\
    OCRVerse~\cite{zhong2026ocrverse} & 2026.1 & 4B & 88.56 & 0.058 & 86.91 & 84.55 & 88.45 & 0.071 \\
    FireRed-OCR~\cite{wu2026firered} & 2026.3 & 2B & 92.94 & 0.032 & 91.71 & 90.31 & 93.81 & 0.041 \\

    \midrule
    \rowcolor[HTML]{EFEFEF} \multicolumn{9}{c}{\textit{Multi-Stage Specialized VLMs}} \\ \midrule
    Dolphin~\cite{feng2025dolphin} & 2025.5 & 0.3M & 74.67 & 0.125 & 67.85 & 68.70 & 77.77 & 0.124 \\
    Dolphin-1.5 ~\cite{dolphin2025} & 2025.5 & 0.3M & 83.21 & 0.092 & 80.78 & 78.06 & 84.10 & 0.080 \\
    MonkeyOCR-pro-1.2B~\cite{li2025monkeyocr} & 2025.6 & 1.2B & 86.96 & 0.084 & 85.02 & 84.24 & 89.02 & 0.130 \\
    MonkeyOCR-3B~\cite{li2025monkeyocr} & 2025.6 & 3B & 87.13 & 0.075 & 87.45 & 81.39 & 85.92 & 0.129 \\
    MonkeyOCR-pro-3B~\cite{li2025monkeyocr} & 2025.6 & 3B & 88.85 & 0.075 & 87.25 & 86.78 & 90.63 & 0.128 \\
    MinerU2.5~\cite{niu2025mineru2} & 2025.10 & 1.2B & 90.67 & 0.047 & 88.46 & 88.22 & 92.38 & 0.044 \\
    PaddleOCR-VL~\cite{cui2025paddleocr} & 2025.10 & 0.9B & 92.86 & 0.035 & 91.22 & 90.89 & 94.76 & 0.043 \\
    PaddleOCR-VL-1.5~\cite{cui2026paddleocr15} & 2026.2 & 0.9B & 94.50 & 0.035 & 94.21 & 92.76 & 95.79 & 0.042 \\ 
    GLM-OCR~\cite{duan2026glm} & 2026.3 & 0.9B & 94.62 & 0.030 & 93.90 & 93.96 & 96.39 & 0.044 \\
    
    \bottomrule
\end{tabular}}
\end{table*}

\subsection{General-Purpose VLMs for Document Parsing}

General-purpose VLMs were not originally designed for structured document parsing; however, they have played a foundational role in advancing multimodal document understanding. Early VLMs already achieved promising results on document parsing tasks. Models such as Qwen2.5-VL-72B~\cite{qwen2.5-VL}, InternVL~\cite{chen2024internvl}, DeepSeek-VL~\cite{lu2024deepseek}, Seed-1.5-VL~\cite{guo2025seed1}, Kimi-VL~\cite{team2025kimi}, and GPT-4o~\cite{hurst2024gpt4o} demonstrated that large transformer-based multimodal architectures pretrained on image--text corpora could handle coarse-grained document tasks, including layout recognition, reading-order reasoning, and element-level question answering. More recent large-scale VLMs, such as InternVL3~\cite{wang2025internvl35}, Qwen3-VL~\cite{qwen3technicalreport}, and Qwen3.5~\cite{Qwen35}, have further advanced these capabilities. Proprietary models, including GPT-5.2~\cite{singh2025gpt5}, Gemini 2.5/3 Pro~\cite{comanici2025gemini}, Kimi-K2.5~\cite{kimik25}, and Claude Sonnet 3/4~\cite{Claude}, have also strengthened cross-modal reasoning and long-context alignment. In zero-shot or instruction-following settings, these models frequently outperform traditional OCR-based pipelines in high-level structural reasoning and semantic interpretation.

The strengths of general-purpose VLMs lie in their scale, broad pretraining, and strong instruction-following capabilities, which enable flexible reasoning over heterogeneous document types without task-specific engineering. However, when applied to fine-grained structured parsing, several intrinsic limitations become apparent. Their training objectives emphasize broad visual grounding and conversational alignment rather than hierarchical structural modeling, leading to strong paragraph-level transcription but unstable page-level organization. Hallucination and repetitive generation are particularly pronounced in dense, professionally typeset documents such as scientific PDFs, suggesting reliance on statistical priors rather than explicit structural constraints. Moreover, the computational overhead associated with multi-billion-parameter models and quadratic attention over high-resolution visual tokens limits their scalability in industrial pipelines. These limitations have motivated the development of document-specialized architectures that embed stronger structural inductive biases.

\subsection{End-to-End Specialized VLMs}

End-to-end document VLMs aim to directly convert raw page images into structured representations such as Markdown, \LaTeX, or HTML. By jointly modeling layout detection, content recognition, and structural relation inference within a unified neural architecture, this paradigm reduces the cascading errors inherent in traditional multi-stage systems. The central hypothesis is that structural coherence is best preserved when layout geometry and textual decoding are optimized simultaneously rather than sequentially.

Early dedicated VLMs such as Nougat~\cite{blecher2023nougat} and mPLUG-DocOwl~\cite{hu2024mplug1.5} demonstrated the feasibility of direct image-to-markup generation. Vary~\cite{wei2025vary} further introduced specialized vision vocabularies to better align dense document regions with structured outputs. Complementary to these architectural advances, Xiao et al.~\cite{xiao2025adaptive} proposed an adaptive markup generation framework that produces structured representations such as Markdown, JSON, and HTML while constructing large-scale datasets (DocMark-Pile and DocMark-Instruct) to improve contextually grounded document understanding and reduce hallucination in complex layouts. GOT~\cite{wei2024general} extended this idea through its ``General OCR Theory'' (OCR-2.0), advocating a unified decoding framework for diverse artificial optical signals—including text, formulas, tables, charts, and even sheet music. By combining a high-compression encoder with a long-context decoder, GOT achieves high-precision full-page parsing at substantially lower inference cost than general multi-billion-parameter VLMs, illustrating how task-aligned architectural design can compensate for brute-force scaling.
These specialized models embed stronger structural priors and are typically more efficient on document-centric tasks than general-purpose VLMs. Their gains are often tied, however, to output formats and training distributions that are narrower than those of broadly pretrained multimodal models.

Subsequent research emphasized structural fidelity, data efficiency, and multilingual generalization. SmolDocling~\cite{nassar2025smoldocling} demonstrated that a compact 256M-parameter model can remain competitive when paired with an appropriate lightweight structured format, highlighting the importance of output representation design. Similarly, UniRec-0.1B~\cite{du2025unirec} explored extreme model efficiency by proposing a unified text-and-formula recognition model with only 0.1B parameters, supported by hierarchical supervision and a semantic-decoupled tokenizer to handle structural variability across document hierarchies. POINTS-Reader~\cite{liu2025points} proposed a distillation-free self-improvement pipeline, combining large-scale synthetic pretraining with iterative real-document filtering, showing that structural consistency can be progressively enhanced without teacher supervision. dots.ocr~\cite{li2025dots} strengthened multilingual robustness by jointly modeling layout and relational structures across 126 languages, reinforcing the role of integrated structural supervision in cross-lingual scenarios.
In contrast to earlier large end-to-end models, this line of work places more emphasis on efficiency and controllability. A limitation is that compact designs and lightweight formats may reduce representational flexibility for highly complex pages.

Nevertheless, purely supervised fine-tuning (SFT) remains fundamentally limited by next-token prediction objectives, which encourage surface-level imitation rather than explicit structural rule acquisition. This often manifests as reading-order inconsistencies in complex layouts, such as multi-column academic papers. To address this, reinforcement learning has been incorporated into end-to-end training. Logics-Parsing~\cite{chen2025logics} introduced a two-stage SFT-then-RL strategy with layout-aware rewards to enforce natural reading sequences. Infinity-Parser~\cite{wang2025infinity} formalized document parsing as a LayoutRL problem, optimizing hierarchical consistency through document-level rewards. olmOCR 2.0~\cite{poznanski2025olmocr} further advanced this direction by introducing Reinforcement Learning with Verifiable Rewards (RLVR), employing binary unit tests to validate mathematical expressions and table structures. By integrating correctness verification into training, these approaches shift optimization from token similarity toward structural validity. Extending this idea toward broader OCR scenarios, OCRVerse~\cite{zhong2026ocrverse} proposes a holistic end-to-end framework that unifies traditional text-centric OCR with vision-centric document understanding tasks such as charts and web pages, combining cross-domain SFT with domain-specific reinforcement learning rewards to support flexible structured outputs across heterogeneous document types.
Relative to standard SFT, these approaches are better aligned with structural correctness rather than surface similarity alone. Their effectiveness, however, depends heavily on reward design, which remains difficult to generalize across heterogeneous document elements.

Architectural innovation has also played a critical role. DeepSeek-OCR~\cite{wei2025deepseek} reframed the vision-language interface from an LLM-centric perspective, treating visual tokens as an efficient compression medium for textual information and achieving 7--20$\times$ effective compression. DeepSeek-OCR 2~\cite{wei2026deepseek2} introduced DeepEncoder V2, replacing CLIP-style encoders with a compact LLM-based architecture that models a ``causal visual flow.'' By reordering visual features according to semantic dependencies rather than fixed raster order, it directly addresses the mismatch between two-dimensional layouts and one-dimensional token sequences.

Recent efforts have also explored systematic approaches to transforming general VLMs into document-specialized models. HunyuanOCR~\cite{team2025hunyuanocr} represents a recent large-scale instantiation of the end-to-end paradigm. With 1B parameters, 200M application-aligned pretraining samples, and online reinforcement learning via GRPO, it achieves strong performance across document parsing, text spotting, and translation. Notably, it demonstrates that task-aligned data curation and reinforcement-based refinement can rival significantly larger general-purpose VLMs.
FireRed-OCR~\cite{wu2026firered}, for example, introduces a progressive training framework that converts a general-purpose VLM into a pixel-precise structural OCR system through geometry-aware data generation and a three-stage training curriculum combining multi-task alignment, structured SFT, and format-constrained GRPO optimization. 
Adaptation-based strategies can reuse general multimodal capabilities more efficiently than training document models from scratch. They may still inherit biases from the source VLM and therefore require careful domain-specific correction.

Overall, end-to-end specialized VLMs substantially enhance structural coherence and simplify system design by unifying layout and recognition. However, requiring a single autoregressive sequence to encode geometry, textual content, and relational dependencies places considerable demands on model capacity and optimization stability. As document layouts grow more complex and dense, scalability and interpretability remain open challenges.

\subsection{Multi-Stage Specialized VLMs}

In response to the scalability and robustness limitations of unified generation, multi-stage or hybrid architectures reintroduce controlled decomposition while retaining neural integration. Instead of collapsing document parsing into a single sequence generation task, these approaches modularize the process into interpretable sub-tasks, thereby improving efficiency and reducing hallucination in high-resolution, text-dense scenarios.

MonkeyOCR~\cite{li2025monkeyocr} introduced the Structure-Recognition-Relation (SRR) paradigm, explicitly separating layout detection, content recognition, and relational modeling. By avoiding full-page quadratic self-attention, it enables block-level parallel recognition while preserving fine-grained details. MonkeyOCR v1.5~\cite{zhang2025monkeyocr} further simplifies the pipeline and incorporates reinforcement learning based on visual consistency, alongside Image-Decoupled Table Parsing (IDTP) for complex and cross-page tables. MinerU 2.5~\cite{niu2025mineru2} proposes a decoupled coarse-to-fine inference mechanism, first performing global layout analysis on downsampled images and subsequently applying high-resolution recognition to cropped regions. This strategy reduces token redundancy and significantly mitigates hallucination, enabling a 1.2B-parameter model to outperform much larger general-purpose VLMs in long-document processing. GLM-OCR~\cite{duan2026glm} exemplifies a compact yet high-performance multi-stage system, integrating a 0.4B-parameter CogViT visual encoder with a 0.5B-parameter GLM language decoder. Using a two-stage pipeline—PP-DocLayout-V3 for layout analysis followed by parallel region-level recognition—and a Multi-Token Prediction mechanism to accelerate decoding, it achieves competitive results in text and formula recognition, table reconstruction, and key information extraction, demonstrating suitability for both resource-constrained edge deployment and large-scale production systems.
These systems generally improve efficiency, controllability, and long-page robustness relative to unified autoregressive generation. The trade-off is that reintroduced decomposition can partially restore inter-stage dependencies that end-to-end models seek to remove.

Hybrid systems also emphasize robustness to real-world distortions. PaddleOCR-VL~\cite{cui2025paddleocr} and its v1.5~\cite{cui2026paddleocr15} upgrade introduce PP-DocLayoutV3, which replaces rectangular bounding boxes with multi-point representations to model non-planar distortions such as warping and skewing. Combined with the Real5-OmniDocBench benchmark, this line of work highlights robustness to physical artifacts as a critical yet underexplored dimension in purely generative models.
This direction better reflects deployment conditions in scanned and photographed documents than benchmarks centered on clean rendered pages. Robustness to physical distortions nevertheless remains uneven across element types and output formats.

In summary, multi-stage and hybrid architectures offer improved computational efficiency, stronger robustness to geometric distortions, and enhanced interpretability through intermediate structural representations. However, reintroducing modular stages increases system complexity and may partially revive inter-stage dependency issues. The ongoing competition between unified end-to-end modeling and structured hybrid decomposition reflects a broader trade-off between architectural elegance and controlled structural precision.

%% file: section/sec10_Discussion.tex
\section{Discussion and Outlook}
\label{sec:discussion}

\subsection{Discussion}
The evolution of document parsing reflects a fundamental shift from modular pipelines to VLM-based unified models. However, both paradigms encounter distinct structural and operational bottlenecks. This section highlights these obstacles and explores potential directions for future research and development.

\paragraph{Pipeline-Based Systems}
Due to their reliance on complex rule-based frameworks, pipeline systems often lack robustness in sophisticated document scenarios. Their primary limitations include:
\begin{enumerate}
    \item \textbf{Error Cascading}: The error tolerance of pipeline structures is relatively low. For instance, minor inaccuracies in layout analysis can lead to severe cascading failures in subsequent OCR or element parsing modules.
    \item \textbf{Hardship in Unified Optimization}: Individual sub-models typically require independent maintenance and training, making it difficult to achieve end-to-end global optimization.
\end{enumerate}

Currently, unified models driven by VLMs are increasingly adopted in place of traditional pipeline-based systems because they can offer stronger end-to-end performance and simplified system design. Nevertheless, enhancing the parsing performance of individual document elements remains indispensable. Precision in parsing single elements---such as dense newspaper layouts, multi-line or handwritten formulas in educational materials, and complex multi-page financial tables---remains highly important. Improving these sub-modules not only provides cost-effective and high-quality solutions for niche applications but also generates high-fidelity synthetic data for training unified models. 

Furthermore, the extraction of visually rich elements (e.g., artistic graphics, natural illustrations, and chemical structures) provides substantial contextual information. As most current models offer limited support for these elements---often merely cropping them from the document---advanced perception and reconstruction of these visual components remain an important direction for future development.

\paragraph{Unified Models Driven by VLM}
Whether general-purpose or specialized for document parsing, VLMs offer a high-precision alternative for structured data extraction. Currently, specialized VLMs follow two primary technical trajectories:
\begin{itemize}
    \item \textbf{Multi-Stage Parsing}: Effectively an advanced version of the pipeline system. While it easily integrates diverse sub-task data, it inevitably faces error propagation and optimization challenges.
    \item \textbf{End-to-End Parsing}: Since the model completes all parsing tasks in a single forward pass, it is more conducive to optimizing overall parsing consistency. From a long-term perspective, end-to-end architectures—integrating native multi-modality and customized vision encoders—possess stronger evolutionary potential.
\end{itemize}

However, the deployment and inference costs associated with VLMs cannot be ignored. Most specialized models therefore focus on maximizing precision while reducing parameter counts and deployment overhead. 

Moreover, the integration of \textbf{reinforcement learning (RL)} in post-training has shown strong potential for mitigating issues such as repetitive generation, grammatical errors, and character confusion. Developing specialized reward models for document parsing is therefore important because it can enable semi-supervised learning on large-scale unlabeled data while helping control data acquisition costs.

\subsection{Outlook}
Future document parsing research, whether modular or unified, must focus on three core dimensions:

\paragraph{Robustness in Complex Scenarios.}
Research should pivot toward real-world challenges, such as blurred, skewed, watermarked, or folded document images. Furthermore, extending support to handwritten documents, historical archives, and low-resource languages will unlock significant latent knowledge value across various industries.

\paragraph{Data Scale and Quality Bottleneck.}
Data remains the primary bottleneck for performance gains. Both technical routes currently rely on extensive supervised fine-tuning (SFT) with high-quality annotated data. However, the diversity of document types is nearly inexhaustible, and large-scale manual annotation is prohibitively expensive. While distilling from closed-source commercial models is a common workaround, it prevents the student model from truly surpassing the teacher's capabilities. Developing low-cost, high-fidelity automated data pipelines is imperative for the next breakthrough.

\paragraph{Application-Oriented Evaluation.}
To continuously enhance parsing capabilities, specialized evaluation frameworks are urgently needed. Although benchmarks like OmniDocBench and OLM-Bench have emerged, they still struggle to cover the infinite variety of documents, and models may inadvertently overfit to these static sets. Since document parsing is a critical upstream component for RAG and AI Agents, its real-world performance directly impacts the quality of knowledge base construction. Developing evaluation schemes that reflect quality in actual industrial application scenarios remains a shared priority for both academia and industry.

%% file: section/sec11_conclusion.tex
\section{Conclusion}
\label{sec:conclusion}

In this survey, we presented a structured and up-to-date overview of document parsing, organizing the field through the lens of both traditional modular pipelines and unified
document parsing models driven by VLMs. We systematically reviewed key components of the parsing workflow, including layout analysis, OCR, mathematical expression recognition, table understanding, and other visual element parsing, along with a comprehensive summary of commonly used datasets and evaluation protocols.
We highlighted the ongoing paradigm shift from task-specific, pipeline-based approaches toward unified, end-to-end multimodal models capable of structured generation. At the same time, we emphasized that modular methods remain essential for achieving robustness and fine-grained accuracy in complex real-world scenarios. 
Looking forward, advancing document parsing will require progress in structure-aware modeling, scalable and high-quality data construction, and more comprehensive evaluation frameworks that better reflect real-world document complexity and multimodal reasoning requirements. 
We hope this survey serves as a coherent reference for researchers and practitioners developing next-generation document intelligence systems.

%% file: section/Appendix.tex
\clearpage
\setcounter{page}{1}

\setcounter{section}{0}
\setcounter{table}{0}
\setcounter{figure}{0}
\renewcommand{\thesection}{\Roman{section}}
\renewcommand{\thetable}{S\arabic{table}}
\renewcommand{\thefigure}{S\arabic{figure}}

\begin{center}
    {\LARGE \textbf{Appendix}}
\end{center}

\section{Evaluation}
\label{sec:eval}

Evaluating document parsing systems requires both fine-grained metrics tailored to heterogeneous subtasks and representative benchmarks that reflect real-world document complexity. This section first reviews commonly adopted metrics across layout, text, formula, and table parsing, and then summarizes task-specific and holistic benchmarks that have shaped the evolution of evaluation protocols. Section~\ref{evaluation_metrics} summarizes representative and widely adopted metrics across these subtasks, while Section~\ref{evaluation_benchmark} introduces benchmarks for both subtask-level and end-to-end document parsing evaluation.

\subsection{Metrics}
\label{evaluation_metrics}

Because document parsing comprises multiple heterogeneous subtasks and its outputs range from bounding boxes to linearized markup and structured trees, evaluation protocols vary accordingly. 

For document layout analysis, evaluation focuses on spatial localization and category prediction of layout elements. The core metric is \emph{Intersection over Union (IoU)}~\cite{rezatofighi2019generalized}:
\begin{equation}
\text{IoU} = \frac{\text{Area of Overlap}}{\text{Area of Union}},
\end{equation}
which measures the overlap between predicted and ground-truth regions. Based on IoU thresholding, \emph{Precision}, \emph{Recall}, and \emph{F1-score} are computed to quantify detection accuracy. To aggregate performance across categories and confidence thresholds, \emph{mean Average Precision (mAP)} is widely adopted:
\begin{equation}
\text{mAP} = \frac{1}{N}\sum_{i=1}^{N} \text{AP}_i,
\end{equation}
where $\text{AP}_i$ denotes the average precision of class $i$. In more comprehensive settings, \emph{mAP@IoU[a:b]} averages mAP over a range of IoU thresholds (e.g., 0.5–0.95), providing a stricter and more robust evaluation of localization quality.
However, fixed IoU thresholding may misalign with perceptual layout quality when annotation granularity differs from reasonable predictions. To alleviate this issue, MinerU 2.5~\cite{niu2025mineru2} proposes \emph{Page-IoU}, a page-level coverage metric that compares predicted and ground-truth layouts via pixel-wise coverage maps over the non-background region $M$:
\begin{equation}
\text{PageIoU}(P,G)
= \frac{\sum_{p\in M} \min\{P_{\text{cover}}(p),\, G_{\text{cover}}(p)\}}
{\sum_{p\in M} \max\{P_{\text{cover}}(p),\, G_{\text{cover}}(p)\}}.
\end{equation}
By aggregating pixel-level minimum and maximum coverage counts instead of enforcing one-to-one box matching, Page-IoU provides a more holistic and granularity-robust assessment of layout consistency.

For text recognition, evaluation is performed at the string level. The most fundamental metric is \emph{Edit Distance} (Levenshtein distance), which measures the minimum number of insertions, deletions, and substitutions required to transform the predicted string into the ground truth. 
Character Error Rate (CER)~\cite{kanai1993performance} and Word Error Rate (WER)~\cite{neudecker2021survey} are normalized forms derived from edit distance. 
In addition, n-gram-based metrics such as \emph{BLEU}~\cite{papineni2002bleu} evaluate partial matching by measuring n-gram overlap between prediction and reference, while \emph{METEOR}~\cite{sun2024locr} incorporates both precision and recall and allows flexible matching through stemming and synonym alignment. These metrics provide complementary perspectives on textual similarity, especially in long-sequence recognition scenarios.

For mathematical expression recognition, evaluation is more challenging due to structural complexity and the existence of multiple equivalent LaTeX representations. Edit-distance-based metrics and \emph{ExpRate} (exact match rate) are commonly used, but they are highly sensitive to syntactic variations. To mitigate representation ambiguity, \emph{Character Detection Matching (CDM)}~\cite{wang2024cdm} evaluates structural alignment at the rendered symbol level:
\begin{equation}
\text{CDM} = \frac{2TP}{2TP + FP + FN}.
\end{equation}
By matching visualized character instances rather than raw LaTeX strings, CDM provides a more structure-aware and representation-invariant evaluation. Other metrics such as MSE or SSIM have been explored by treating rendered formulas as images, but they are less frequently adopted in practice.

For table recognition, evaluation must account for both structural topology and cell content. When tables are serialized into structured markup (e.g., HTML or LaTeX), edit distance is often used as a simple similarity measure. However, string-level matching cannot explicitly capture hierarchical structure. To address this limitation, \emph{Tree-Edit-Distance-based Similarity (TEDS)}~\cite{zhong2020image} measures similarity by computing the normalized tree edit distance between predicted and ground-truth HTML trees:
\begin{equation}
\text{TEDS}(T_1,T_2)=1-\frac{\text{TED}(T_1,T_2)}{\max(\text{size}(T_1),\text{size}(T_2))}.
\end{equation}
TEDS jointly considers structural tags and cell content, enabling unified structural–semantic evaluation. \emph{S-TEDS}~\cite{huang2023improving} further simplifies TEDS by ignoring cell content and focusing solely on logical structure (row, column, and spanning relations), making it suitable for structure-centric benchmarking. Other fine-grained metrics, such as multi-column recall (MCR) and multi-row recall (MRR)~\cite{kayal2023tables}, have also been proposed for specific structural aspects.

For chart-related tasks, evaluation protocols depend on task formulation. Element detection commonly adopts IoU-based Precision, Recall, F1-score, and mAP~\cite{ma2021towards}. However, for data extraction tasks (e.g., recovering numeric values or keypoints), detection-style metrics are insufficient. In line-chart extraction, Object Keypoint Similarity (OKS) and its strict/relaxed variants measure geometric deviation between predicted and ground-truth keypoints under scale normalization, directly reflecting numerical recovery accuracy. StructChart~\cite{xia2023structchart} proposes the Structuring Chart-oriented Representation Metric (SCRM), which computes precision under fixed IoU thresholds and averages performance across multiple thresholds, resembling an mAP-style aggregation tailored to structured chart representation. Overall, chart evaluation remains less standardized than layout or OCR tasks and is often task-specific.

It can be observed that edit-distance-based metrics are widely used across recognition-oriented subtasks in document parsing. In many works, model outputs are first normalized into a unified representation (e.g., converting tables into HTML) or cleaned through rule-based preprocessing (e.g., removing formatting control symbols), after which edit distance is computed as a document-level score. While convenient and general, such holistic string-based evaluation has notable limitations.

OmniDocBench~\cite{ouyang2024omnidocbench} addresses this issue by proposing a unified and multi-dimensional evaluation framework that supports end-to-end, task-specific, and attribute-based assessment, enabling fine-grained analysis across document types and element categories rather than relying solely on global text similarity. CC-OCR~\cite{yang2025cc} further designs a comprehensive benchmark tailored to large multimodal models, covering multiple OCR-centric tasks and structured outputs, thereby highlighting the necessity of task-aware and challenge-oriented evaluation beyond fragmented single-task metrics. In contrast to continuous edit-distance scoring, olmOCR-Bench~\cite{poznanski2025olmocr} introduces a binary \emph{unit-test-based} evaluation protocol, where predictions are validated against executable test cases (e.g., text presence, reading order, table structure, and formula rendering correctness). This design allows equivalently correct representations to receive consistent scores and better aligns evaluation results with practical correctness in downstream usage.

In summary, although string-based similarity metrics remain prevalent due to their simplicity and universality, recent benchmarks increasingly emphasize structure-aware, task-decomposed, and functionality-oriented evaluation protocols to better reflect real-world document parsing performance.

\subsection{Benchmarks}
\label{evaluation_benchmark}

Benchmarks for document parsing have evolved alongside modeling paradigms, progressing from task-isolated datasets to unified and robustness-oriented evaluation suites, as summarized in Table~\ref{tab:doc_parsing_benchmark_comparison}.

\begin{table}[htbp]
\centering
\small
\setlength{\tabcolsep}{5pt}
\begin{adjustbox}{max width=\textwidth} 
    \begin{tabular}{l l c c c c}
    \toprule
    \textbf{Tasks} & \textbf{Benchmark} & \textbf{Domains} & \textbf{Test Samples} & \textbf{Languages} & \textbf{Type} \\
    \midrule
    
    \multirow{4}{*}{DLA} 
    & DocBank~\cite{li2020docbank} 
    & 1 (academic) & 400,000 & EN & printed \\
    & D$^4$LA~\cite{da2023vision} 
    & 1 (mixed noisy) & 2,224 & EN & printed \\
    & DocLayNet~\cite{pfitzmann2022doclaynet} 
    & 6 & 80863(train/test/val) & Multi & printed \\
    & M$^6$Doc~\cite{cheng2023m6doc} 
    & 6 & 9080(train/test/val) & Multi & scanned/printed/photographed \\
    
    \midrule
    
    \multirow{2}{*}{MER} 
    & CROHME~\cite{mouchere2014icfhr,mouchere2016icfhr2016,mahdavi2019icdar} 
    & 1 & $\sim$3K (test) & EN & handwritten \\
    & UniMER-Test~\cite{wang2024unimernet} 
    & 4 sub-tasks & 23,757 & EN & printed/handwritten \\
    
    \midrule
    
    \multirow{2}{*}{TR} 
    & PubTabNet~\cite{smock2022pubtables} 
    & 1 (scientific) & 9K & EN & printed \\
    & FinTabNet~\cite{zheng2021global} 
    & 1 (financial) & 10K & EN & printed \\
    
    \midrule
    
    \multirow{7}{*}{DP} 
    & OmniDocBench~\cite{ouyang2024omnidocbench} 
    & 9 & 981 & Multi & printed/scanned/handwritten \\
    & OmniDocBench v1.5~\cite{ouyang2024omnidocbench} 
    & 9 & 1355 & Multi & printed/scanned/handwritten \\
    & Real5-OmniDocBench~\cite{zhou2026real5} 
    & 9 & 1355 & Multi & 5 real-world scenarios \\
    & CC-OCR~\cite{yang2025cc} 
    & 39 sub-tasks & 7,058 & Multi & printed/scanned/photographed/handwritten \\
    & OCRBench v2~\cite{fu2024ocrbench}
    & 31 sub-tasks & 1,500 & Multi & printed/scanned/photographed/handwritten \\
    & OceanOCR~\cite{chen2025ocean}
    & Multi & 100 & Multi & scanned/handwritten \\
    & olmOCR-Bench~\cite{poznanski2025olmocr} 
    & Multi & 1402 & Multi & printed \\
    & DocPTBench~\cite{du2025docptbench}
    & Multi-domain & 1300 & Multi & printed/scanned \\
    & Real5-OmniDocBench~\cite{cui2026paddleocr15}
    & 5 perturbation types & 1355 & Multi & scanned/printed \\
    
    \bottomrule
    \end{tabular}
    \end{adjustbox}
    \caption{Comparison of representative document parsing benchmarks across subtasks and holistic evaluation settings. ``Domains'' refers to distinct document types or scenarios covered, and ``Type'' describes the primary document acquisition format included in the benchmark. Statistics are summarized from publicly reported descriptions.}
\vspace{-5mm}
\label{tab:doc_parsing_benchmark_comparison}
\end{table}

Before the widespread adoption of end-to-end and VLM-based systems, research primarily focused on improving individual subtasks within pipeline architectures, leading to a series of high-quality task-specific benchmarks. For document layout analysis, representative datasets include D$^4$LA~\cite{da2023vision}, which contains 11,092 noisy document images annotated with 27 layout categories, and DocLayNet~\cite{pfitzmann2022doclaynet}, a large-scale benchmark with 80,863 manually annotated pages spanning seven document types. Additional datasets such as DocBank~\cite{li2020docbank} and M$^6$Doc~\cite{cheng2023m6doc} provide fine-grained token- or region-level annotations for layout modeling. In mathematical expression recognition, CROHME~\cite{mouchere2014icfhr, mouchere2016icfhr2016, mahdavi2019icdar} has long served as the standard for handwritten formulas, while UniMER-Test~\cite{wang2024unimernet} targets diverse real-world printed and handwritten equation scenarios. For table recognition, PubTabNet~\cite{smock2022pubtables} established large-scale HTML-based supervision for scientific tables, and FinTabNet~\cite{zheng2021global} introduced financial tables with domain-specific structural characteristics. These benchmarks enabled systematic comparison within individual subtasks and drove steady performance improvements under well-defined settings.

With the transition from pipeline systems to end-to-end multimodal models, evaluation increasingly shifted toward holistic document parsing. Early efforts, such as the evaluation suite used in GOT~\cite{wei2024general}, covered multiple OCR-centric document understanding tasks (e.g., plain OCR, formatted document OCR, referential OCR) and relied primarily on string-based similarity metrics. Although diversified in scenario design, such benchmarks largely measured page-level textual overlap and were limited in structural analysis. OmniDocBench~\cite{ouyang2024omnidocbench} represents a more comprehensive attempt, integrating nine document types with multi-level annotations to support end-to-end, task-specific, and attribute-based evaluation across layout, reading order, OCR, tables, and formulas within a unified framework.

Recent benchmarks further expand evaluation toward multimodal literacy and robustness. CC-OCR~\cite{yang2025cc} and OCRBench v2~\cite{fu2024ocrbench} organize OCR-centric capabilities into structured task groups covering multilingual reading, layout-aware parsing, grounding, and key information extraction. Beyond assessing the visual recognition ability of large multimodal models, they provide challenging structured document samples—particularly in complex tables and layout-aware reading—that serve as valuable evaluation resources for document parsing research. In parallel, OceanOCR~\cite{chen2025ocean} emphasizes real-world conditions such as dense bilingual documents and handwriting, highlighting performance gaps under practical deployment scenarios. 

More recently, evaluation has moved toward interpretation-aware and robustness-oriented frameworks. SCORE analyzes the mismatch between deterministic exact-match metrics and generative document representations, advocating disentangled assessment of content fidelity, hallucination, structural consistency, and table semantics to reduce bias caused by representational variance. DocPTBench~\cite{du2025docptbench} introduces over 1,300 human-annotated photographed documents and supports joint evaluation of structured parsing and document translation, exposing substantial degradation of both specialized systems and MLLMs under real capture conditions. Building upon OmniDocBench, PaddleOCR-VL-1.5~\cite{cui2026paddleocr15} constructs Real5-OmniDocBench~\cite{zhou2026real5} to simulate five realistic disturbances—scanning, warping, screen photography, illumination variation, and skew—while preserving one-to-one ground-truth correspondence, enabling controlled robustness evaluation. 

In addition to traditional ground-truth matching protocols, olmOCR-Bench~\cite{poznanski2025olmocr} questions continuous edit-distance scoring and proposes executable unit-test-based validation for properties such as reading order correctness, table structure integrity, and formula rendering consistency. 

Overall, document parsing benchmarks have progressed from isolated, subtask-driven datasets toward unified, multimodal, and robustness-aware evaluation frameworks. This trajectory reflects a growing recognition that accurate document parsing must be assessed not only by surface-level string similarity, but also by structural fidelity, interpretative flexibility, and resilience under real-world conditions.

\section{Datasets}
\label{sec:datasets}
\subsection{Datasets for DLA}
% Datasets for DLA are primarily classified into synthetic, real-world (Documents and scanned images), and hybrid datasets. Early efforts focused on historical documents, after 2010, research interest has transitioned towards complex printed layouts alongside the continued examination of handwritten historical texts. Table~\ref{tab:DLA} lists key datasets used in DLA research over the last ten years.
Early work on datasets for DLA tasks focused on historical documents, such as IMPACT ~\cite{papadopoulos2013impact}, Saint Gall ~\cite{binmakhashen2019document}, and GW20 ~\cite{krishnamoorthy1993syntactic}. More comprehensive datasets have emerged, such as IIT-CDIP ~\cite{lewis2006building}, which contains 7 million documents with complex layouts. After 2010, research interest shifted to complex typographic layouts~\cite{cheng2023m6doc, da2023vision, zhao2024doclayout}, while continuing to study handwritten historical texts. Table~\ref{tab:DLA} lists the key datasets used in DLA research over the past decade.
% In addition, competitions held by major conferences such as the International Conference on Document Analysis and Recognition (ICDAR) have introduced datasets with high-quality, standardized annotations. These are essential for model evaluation and benchmarking. For example, the ICDAR 2013 page segmentation competition focused on document layout analysis using newspapers, journals, and magazines with multiple annotation types. The ICDAR 2021 competition emphasizes historical documents, addressing layout challenges due to aging, and scientific document parsing for extracting structured information.

\begin{table}[h]
    \caption{A detailed list of datasets for document layout analysis.}
    \vspace{-3mm}
    \centering
    \small 
    \setlength{\tabcolsep}{4pt} 
    \renewcommand{\arraystretch}{1} 
    \begin{tabular}{@{} l c c p{4cm} l @{}}
        \toprule
        \textbf{Dataset} & \textbf{Class} & \textbf{Instance} & \textbf{Document Type} & \textbf{Language} \\ \midrule
        PRImA ~\cite{antonacopoulos2009realistic}& 10 & 305 & Multiple Types & English \\
        BCE-Atabic-v1~\cite{saad2016bce} & 3 & 1833 & Arabic books & Arabic \\
        Diva-hisdb~\cite{simistira2017icdar2017} & Text Block& 150 & Handwritten Historical Document & Multiple Languages \\
        DSSE200~\cite{yang2017learning} & 6 & 200 & Magazines, Academic papers & English \\
        OHG~\cite{quiros2018multi} & 6 & 596 & Handwritten Historical Document & English \\
        CORD~\cite{park2019cord} & 5 & 1000 & Receipts & Indonesian \\
        FUNSD~\cite{sahu2017study} & 4 & 199 & Form document & English \\
        PubLayNet~\cite{zhong2019publaynet} & 5 & 360000 & Academic papers & English \\
        Chn~\cite{li2020cross} & 5 & 8005 & Chinese Wikipedia pages & Chinese \\
        DocBank~\cite{li2020docbank}& 13 & 500000 & Academic papers & English, Chinese \\
        BCE-Arabic~\cite{elanwar2021extracting} & 21 & 9000 & Arabic books & Arabic \\
        DAD~\cite{markewich2022segmentation} & 5 & 5980 & Articles & English \\
        DocLayNet~\cite{pfitzmann2206doclaynet} & 11 & 80863 & Multiple Types & Primarily English \\
        D4LA~\cite{da2023vision} & 27 & 11092 & Multiple Types & English \\
        M6Doc~\cite{cheng2023m6doc}& 74 & 9080 & Multiple Types & English, Chinese \\
        DocSynth-300K~\cite{zhao2024doclayout}& 74 & 300,000 & Multiple Types & English, Chinese \\
        \bottomrule
    \end{tabular}
    \label{tab:DLA}
    \vspace{-5mm}
\end{table}

\subsection{Datasets for OCR}
% In terms of OCR datasets, scene text OCR datasets still dominate, and also contain a large amount of artificially synthesized data. There are also some works that have compiled datasets related to text recognition in documents, as shown in Table~\ref{tab:OCR}.
In terms of OCR datasets for printed text, the most notable and widely used datasets are those introduced in various ICDAR competitions, such as ICDAR2013~\cite{gobel2013icdar} and ICDAR2015~\cite{karatzas2015icdar}, which include real-world scenes and document images and are often used to evaluate scene text detection algorithms. In addition, datasets such as Street View Text Perspective and MSRA-TD500~\cite{yao2012detecting} focus on detecting irregular text in challenging environments. Synthetic datasets such as SynthText~\cite{gupta2016synthetic} and SynthAdd~\cite{litman2020scatter} are artificially generated and contain a large amount of data, making them popular in text detection and recognition tasks. In addition, datasets such as ICDAR2019~\cite{gao2019icdar} provide region annotations and text content, supporting end-to-end OCR tasks. A summary of commonly used OCR datasets is provided in Table~\ref{tab:OCR}.

\begin{table}[h]
    \caption{A detailed list of datasets for optical character recognition.}
    \vspace{-3mm}
    \centering
    \small
    \setlength{\tabcolsep}{4pt} 
    \renewcommand{\arraystretch}{1} 
    \begin{tabular}{@{} c c c c c @{}}
        \toprule
        \textbf{Dataset} & \textbf{Instance} & \textbf{Task} & \textbf{Type} & \textbf{Language} \\ \midrule
        IIIT5K~\cite{mishra2012scene} & 5000 & TR & Real-world scene text & En \\
        Street View Text~\cite{jaderberg2016reading} & 647 & TD & Street View & En \\
        Street View Text Perspective~\cite{shi2016robust} & 645 & TD & Street view with perspective distortion & En \\
        ICDAR 2003~\cite{lucas2005icdar} & 507 & TD \& TR & Real-world short scene text & En \\
        ICDAR 2013~\cite{karatzas2013icdar}& 462 & TD \& TR & Real-world short scene text & En \\
        MSRA-TD500~\cite{yao2012detecting} & 500 & TD & Rotated text & En, Zh \\
        CUTE80~\cite{risnumawan2014robust} & 13000 & TD \& TR & Curved text & En \\
        COCO-Text~\cite{veit2016coco} & 63,686 & TD \& TR & Real-world short scene text & En \\
        ICDAR 2015 ~\cite{karatzas2015icdar} & 1670 & TD \& TR \& TS & Scene text and video text & En \\
        SCUT-CTW1500~\cite{liu2019curved} & 1500 & TD & Curved text & En, Zh \\
        Total-Text~\cite{ch2017total} & 1555 & TD \& TR & Multi-oriented scene text & En, Zh \\
        SynthText~\cite{gupta2016synthetic} & 800,000 & TD \& TR & Synthetic images & En \\
        SynthAdd~\cite{litman2020scatter} & 1,200,000 & TD \& TR & Synthetic images & En \\
        Occlusion Scene Text ~\cite{wang2021two} & 4832 & TD & Occlusion text & En \\
        WordArt~\cite{xie2022toward} & 6316 & TR & Artistic text & En \\
        ICDAR2019-ReCTS ~\cite{zhang2019icdar} & 25,000 & TD \& TR \& TS & Chinese signboards & Zh \\
        LOCR ~\cite{sun2024locr} & 7,000,000 & TD \& TR \& TS & Academic text & Zh \\
        \bottomrule
        \multicolumn{4}{l}{\small TD: Text Detection; TR: Text Recognition; TS: Text Spotting.}\\
    \end{tabular}
    \vspace{-6mm}
    \label{tab:OCR}
\end{table}
\subsection{Datasets for MED and MER}
% In document analysis, mathematical expression detection and recognition are crucial research areas. With specialized datasets, researchers now achieve improved recognition of diverse mathematical mathematical expressions. Table~\ref{tab:MED-MER} lists common benchmark datasets for mathematical expression detection and recognition, covering both printed and handwritten mathematical expressions across various document formats like images and Documents. These datasets support tasks such as mathematical expression detection, extraction, localization, and mathematical expression recognition.

For tasks such as math expression detection, extraction, localization, and math expression recognition, important datasets include UW-III~\cite{liang1997performance}, InftyCDB-1~\cite{suzuki2005ground}, and Marmot~\cite{lin2012performance}, which are commonly used to evaluate printed math expression detection, solving inline and standalone math expressions. The ICDAR series has promoted the development of this field through competitions on datasets such as ICDAR-2017 POD~\cite{gao2017icdar2017} and ICDAR-2021 IBEM~\cite{anitei2021icdar}, showing a wide range of complex scenarios. These resources improve the robustness of recognition models and highlight the challenges of detecting math expressions in complex document structures. In addition, datasets such as FormulaNet~\cite{schmitt2022formulanet} and ArxivFormula~\cite{hu2024mathematical} emphasize large-scale detection, especially extracting math expressions from images~\ref{tab:MED-MER} . Despite the progress, datasets for math expression detection and recognition are still limited, so there is a need to improve multi-format support and robustness.

\begin{table}[h]
    \caption{A detailed list of datasets for mathematical expression detection and recognition.}
    \centering
    \setlength{\tabcolsep}{3pt} 
    \renewcommand{\arraystretch}{1} 
    \begin{adjustbox}{max width=\textwidth} 
    \begin{tabular}{@{} l c c c l @{}}
        \toprule
        \textbf{Dataset} & \textbf{Image} & \textbf{Instance} & \textbf{Type} & \textbf{Task} \\ \midrule
        UW-III~\cite{liang1997performance} & 100 & / & Inline and displayed Formula & MED \\
        InftyCDB-1~\cite{suzuki2005ground} & 467 & 21,000 & Inline and displayed Formula & MED \\
        Marmot~\cite{lin2012performance} & 594  & 9,500 & Inline and displayed Formula & MED \\
        ICDAR-2017 POD~\cite{gao2017icdar2017} & 3,900 & 5,400 & Only displayed Formula & MED \\
        TFD-ICDAR 2019 ~\cite{mahdavi2019icdar} & 851 & 38,000 & Inline and displayed Formula & MED \\
        ICDAR-2021 IBEM~\cite{anitei2021icdar} & 8,900 & 166,000 & Inline and displayed Formula & MED \\
        FormulaNet~\cite{schmitt2022formulanet} & 46,672 & 1,000,00 & Inline and displayed Formula & MED \\
        ArxivFormula~\cite{hu2024mathematical} & 700,000 & 813.3 & Inline and displayed Formula & MED \\
        Im2Latex-100K~\cite{deng2017image} & \multicolumn{2}{c}{\centering 103,556} & Printed & MER \\
        Pix2tex~\cite{pix2tex2022} & \multicolumn{2}{c}{\centering 189117} & Printed & MER \\
        CROHME~\cite{mouchere2014icfhr} & \multicolumn{2}{c}{\centering 12178} & Handwritten & MER \\
        HME100K~\cite{yuan2022syntax} & \multicolumn{2}{c}{\centering 99109} & Handwritten & MER \\
        UniMER-1M~\cite{wang2024unimernet} & \multicolumn{2}{c}{\centering 1,061,791} & Printed and Handwritten & MER \\
        \bottomrule
    \end{tabular}
    \end{adjustbox}
    \label{tab:MED-MER}
\end{table}

\subsection{Datasets for TD and TR}
Tabular data is diverse and complex in structure, and a large number of representative datasets have emerged in table-related tasks. Basic and widely applicable table datasets mainly come from the ICDAR official competition~\cite{gobel2013icdar,gao2019icdar,davila2019icdar,anitei2021icdar}. 
Some datasets are specifically for irregular table samples, such as the Marmot~\cite{fang2012dataset} dataset that focuses on the detection of wired and wireless tables, CamCap~\cite{seo2015junction} collects irregular tables photographed on curved surfaces, and the Wired Table in the Wild (WTW)~\cite{long2021parsing} dataset includes common challenging tables in real life such as occlusion and blur. 
These datasets improve the robustness of table recognition systems in complex environments.
Some datasets are also tailored for specific table-related tasks~\cite{chi2019complicated, smock2022pubtables,nassar2022tableformer,li2021gfte,desai2021tablex,zheng2021global}.
For example, FinTabNet~\cite{zheng2021global} focuses on the detection and recognition of financial tables and SciTSR~\cite{chi2019complicated} focuses on the recognition of table structures in academic articles. These datasets provide targeted support for professional table analysis tasks and promote progress in segmented research fields. In addition, datasets such as WikiTableSet~\cite{ly2023rethinking} covers tables in multiple languages, including Chinese, which helps solve the problem of insufficient language diversity and realize cross-language table detection and structural analysis.
The datasets for table detection and table structure recognition are summarized in Table \ref{tab:TD-TSR}.
Although existing table datasets provide rich data sources and diverse scenario support for tasks such as table detection and structural recognition, there is still room for improvement in terms of scenario diversity, task targeting, and language coverage.

\begin{table}[t]
    \caption{A detailed list of datasets for table detection and structure recognition.}
    \vspace{-3mm}
    \centering
    \small
    \setlength{\tabcolsep}{3pt} 
    \renewcommand{\arraystretch}{1} 
    \begin{adjustbox}{max width=\textwidth} 
    \begin{tabular}{@{} l c c c c @{}}
        \toprule
        \textbf{Dataset} & \textbf{Instance} & \textbf{Task} & \textbf{Type} & \textbf{Language} \\ \midrule
        ICDAR2013 ~\cite{gobel2013icdar} & 150 & TD \& TSR & Government Documents & En \\
        ICDAR2017 POD~\cite{gao2017icdar2017} & 1548 & TD & Academic papers & En \\
        ICDAR2019 ~\cite{gao2019icdar}& 2439 & TD \& TSR & Multiple Types & En \\
        TABLE2LATEX-450K ~\cite{deng2019challenges} & 140000 & TSR & Academic papers & En \\
        RVL-CDIP (subset) ~\cite{riba2019table} & 518 & TD & Receipts & En \\
        IIIT-AR-13K~\cite{mondal2020iiit} & 17,000 & TD & Annual Reports & Multi \\
        CamCap~\cite{seo2015junction} & 85 & TD \& TSR & Table images & En \\
        UNLV Table~\cite{shahab2010open} & 2889 & TD & Journals, Newspapers, Business Letters & En \\
        UW-3 Table~\cite{phillips1996user} & 120 & TD & Books, Magazines & En \\
        Marmot~\cite{fang2012dataset} & 2000 & TD & Conference Papers & En, Zh \\
        TableBank~\cite{li2020tablebank} & 417234 & TD \& TSR & Multiple Types & En \\
        DeepFigures~\cite{siegel2018extracting} & 5,500,000 & TD & Academic papers & En \\
        PubTabNet~\cite{zhong2020image} & 568000 & TSR & Academic papers & En \\
        PubTables-1M ~\cite{smock2022pubtables} & 1000000 & TSR & Academic papers & En \\
        SciTSR~\cite{chi2019complicated} & 15000 & TSR & Academic papers & En \\
        FinTable~\cite{zheng2021global} & 112887 & TD \& TSR & Academic and Financial Tables & En \\
        SynthTabNet ~\cite{nassar2022tableformer} & 600000 & TD \& TSR & Multiple Types & En \\
        Wired Table in the Wild ~\cite{long2021parsing} & 14582 pages & TSR & Photos, Files, and Web Pages & En \\
        WikiTableSet~\cite{ly2023rethinking} & 50000000 & TSR & Wikipedia & Multi \\
        STDW~\cite{haloi2022table} & 7000 & TD & Multiple Types & En \\
        TableGraph-350K~\cite{xue2021tgrnet} & 358,767 & TSR & Academic Table & En \\
        TabRecSet~\cite{yang2023large} & 38100 & TSR & Multiple Types & En, Zh \\
        DECO~\cite{koci2019deco} & 1165 & TD & Multiple Types & En \\
        iFLYTAB~\cite{zhang2024semv2} & 17291 & TD \& TSR & Multiple Types & En, Zh \\
        FinTab ~\cite{li2021gfte} & 1,600 & TSR & Financial Table & Zh \\
        TableX ~\cite{desai2021tablex} & 4,000,000 & TSR & Academic papers & En \\
        \bottomrule
        \multicolumn{5}{l}{\small TD: Table Detection; TSR: Table Structure Recognition}\\
    \end{tabular}
    \end{adjustbox}
    \label{tab:TD-TSR}
    \vspace{-5mm}
\end{table}